\documentclass[prd,twocolumn,nofootinbib,superscriptaddress,nolongbibliography]{revtex4-1} 

\usepackage{graphicx,amscd,amsmath,amssymb,verbatim,color}
\usepackage{hyperref}
\usepackage[caption=false]{subfig}

\newcommand{\mnras}{Mon. Not. R. Astron. Soc.}

\begin{document}

\title{21\nobreakdash-cm Constraints on Dark Matter Annihilation after an Early Matter-Dominated Era}
\author{Hwan Bae}
\affiliation{Department of Physics and Astronomy, University of North Carolina at Chapel Hill, Phillips Hall CB3255, Chapel Hill, North Carolina 27599, USA}
\author{Adrienne L. Erickcek}
\affiliation{Department of Physics and Astronomy, University of North Carolina at Chapel Hill, Phillips Hall CB3255, Chapel Hill, North Carolina 27599, USA}
\author{M. Sten Delos}
\affiliation{Carnegie Observatories, 813 Santa Barbara Street, Pasadena, CA 91101, USA} 
\author{Julian B.~Mu\~noz}
\affiliation{Department of Astronomy, The University of Texas at Austin, 2515 Speedway, Stop C1400, Austin, TX 78712, USA} 

\date{\today}

\begin{abstract}
Although it is commonly assumed that relativistic particles dominate the energy density of the universe quickly after inflation, a variety of well-motivated scenarios predict an early matter-dominated era (EMDE) before the onset of Big Bang nucleosynthesis. Subhorizon dark matter density perturbations grow faster during an EMDE than during a radiation-dominated era, leading to the formation of ``microhalos" far earlier than in standard models of structure formation. This enhancement of small-scale structure boosts the dark-matter annihilation rate, which contributes to the heating of the intergalactic medium (IGM).
We compute how the dark matter annihilation rate evolves after an EMDE and forecast how well measurements of the 21\nobreakdash-cm background can detect dark matter annihilation in cosmologies with EMDEs.
We find that future measurements of the global 21\nobreakdash-cm signal at a redshift of $z\sim 17$ are unlikely to improve on bounds derived from observations of the isotropic gamma-ray background, but measurements of the 21\nobreakdash-cm power spectrum have the potential to detect dark matter annihilation following an EMDE.
Moreover, dark matter annihilation and astrophysical X rays produce distinct heating signatures in the 21\nobreakdash-cm power spectrum at redshifts around 14, potentially allowing differentiation between these two IGM heating mechanisms.
\end{abstract}

\maketitle

\section{Introduction}\label{section:Introduction}


Observations of the 21\nobreakdash-cm line, which arises from the transition between the two hyperfine levels of neutral hydrogen, promise to provide new insights into the period between recombination and reionization~\cite{Furlanetto2006,Pritchard2011}. Radio interferometers such as HERA~\cite{HERACollaboration2021}, LOFAR~\cite{Mertens2020}, SKA~\cite{Koopmans2015, Vrbanec2020}, and MWA~\cite{Trott2021, Rahimi2021} measure anisotropies in the 21\nobreakdash-cm background, while the sky-averaged (global) 21\nobreakdash-cm signal can be measured using a small number of broad-beam, wide-band antennae~\cite{Anstey2021}. Several global 21\nobreakdash-cm signal experiments are underway, including EDGES~\cite{Bowman2018}, SARAS~\cite{Nambissan2021,Bevins2022}, SCI-HI~\cite{Voytek2013, Peterson2014}, and LEDA~\cite{Price2017}, and other experiments (such as BIGHORNS~\cite{Sokolowski2015}, PRIZM~\cite{Philip2018}, and REACH~\cite{DeLeraAcedo2019}) are in different stages of planning and construction.

The 21\nobreakdash-cm signal measures the absorption or emission of 21\nobreakdash-cm wavelength photons with respect to that expected given the temperature of the cosmic microwave background ($T_\text{CMB}$). The observed $\delta T_{21}$ is proportional to the difference between the spin temperature $T_S$ and $T_\text{CMB}$, where $T_S$ determines the relative population of the two hydrogen spin states~\cite{Pritchard2011}. Atomic collisions and scattering of Lyman-$\alpha$ photons~\cite{Wouthuysen1952,Field1958} couple the spin temperature to the kinetic temperature of the intergalactic medium (IGM). Consequently, the global 21\nobreakdash-cm signal provides a direct probe of the IGM temperature $T_K$~\cite{Furlanetto2006}.
After recombination, $T_K$ is coupled to the CMB temperature through Compton scattering until a redshift of $z\approx 150$. The IGM temperature then cools as $(1+z)^2$ via adiabatic expansion until heating due to stellar sources becomes significant~\cite{Pritchard2011}. A deviation from this standard evolution would be visible in the 21\nobreakdash-cm global signal~\cite{AliHaimoud2010, Chluba2011} and would suggest the presence of additional heating or cooling, including high-energy particles resulting from dark matter (DM) annihilation~\cite{Chen2004, Furlanetto2006_dmOn21cmBG, Valdes2007, Belikov2009, Slatyer2009,Liu2018,Qin:2023kkk}, as we will study here.

Given standard assumptions about structure formation, the effect of annihilating DM on the IGM temperature is negligible compared to the effect of astrophysical sources such as active galactic nuclei or the first stars~\cite{Araya2013,Poulin2015}.
However, there are a number of early-universe scenarios that enhance structure growth, such as a period of kination~\cite{Redmond2018}, the presence of primordial black holes~\cite{Kadota2021}, or an early matter-dominated era (EMDE)~\cite{Erickcek2011}. Although the standard cosmological model assumes that the universe was radiation dominated between inflation and the onset of matter domination at $z_\mathrm{eq}\simeq 3400$~\cite{Planck2018_VI},
many well-motivated scenarios predict an EMDE during the first second after inflation~\cite{Kane2015, Zhang2015, Allahverdi2021}. This EMDE could be driven by heavy particles or oscillating scalar fields. Any such particle or field must decay into Standard Model (SM) particles prior to neutrino decoupling to ensure that it does not affect light-element abundances~\cite{Kawasaki1999,Hannestad2004,Ichikawa2005} or the neutrino density measured from the CMB~\cite{Ichikawa2007,DeBernardis2008}.

Since subhorizon DM density perturbations grow linearly with the scale factor during the EMDE~\cite{Erickcek2011,Barenboim2013,Fan2014,Erickcek2015}, EMDE cosmologies generate microhalos that form earlier and are denser than those in standard cosmology~\cite{Erickcek2011,Barenboim2013,Erickcek2015,Blanco2019}. These substructures boost the DM annihilation rate, and the resulting energy injection can dramatically alter the evolution of $T_K$. Therefore, an upper bound on $T_K$  places constraints on the DM annihilation rate, as excessive energy injection could heat the IGM temperature beyond this maximum threshold. 
The 21\nobreakdash-cm line can provide measurements of $T_K$ during cosmic dawn, opening a window to test these models.

Global 21\nobreakdash-cm experiments are beginning to report bounds on the differential brightness temperature $\delta T_{21}$. LEDA reported that \mbox{$\delta T_{21} > -890$~mK} in 50-87 MHz with 95\% CL. The EDGES Collaboration reported a value of  \mbox{$ -1000$~mK $ \leq \delta T_{21} \leq -300$~mK}
at $\approx 78$~MHz ($z\approx 17.2$) with 99\%~CL. The EDGES result corresponds to IGM temperatures between 1.7~K and 5.1~K. However, at \mbox{$z\approx 17$}, the standard expectation is that $\delta T_{21} \approx -200$~mK, corresponding to $T_K \approx 7$~K. Scenarios proposed to explain this discrepancy include mechanisms to cool the gas further than allowed by adiabatic expansion~\cite{Barkana2018,Munoz2018,Hill2018,Falkowski2018,Munoz:2018jwq} and incorporating additional radio background sources other than the CMB~\cite{Feng2018,Fialkov2019}. 
Other possibilities include systematics~\cite{Bradley2018}, as well as analysis methods~\cite{Hills2018}, and a recent nondetection from the SARAS3 experiment has cut into the parameter space preferred by EDGES~\cite{Bevins2022}.

The IGM heating from DM annihilation suppresses the observed absorption signal and can even erase the signal altogether. The impact of DM annihilation on the thermal history of the IGM~\cite{Araya2013,Poulin2015,Liu2016,Liu2019} and the resulting 21\nobreakdash-cm signal has been extensively studied~\cite{Furlanetto2006_dmOn21cmBG,Valdes2007,Evoli2014,Poulin2017,Kaurov2016}. There have also been forecasts of the bounds on the velocity-averaged cross section of DM annihilation, $\langle \sigma v \rangle$, that would result if an absorption signal persisted at $z = 17$~\cite{DAmico2018,Liu2018}. However, these forecasts assume standard structure formation scenarios.  To determine how an EMDE boosts DM annihilation, we use the Peak-to-Halo (P2H)~\cite{Delos2019_PredictDens} method to compute the dark matter annihilation rate within the microhalos that form following an EMDE. We use the resulting boost factors to forecast the bounds on the velocity-averaged DM annihilation cross section $\langle \sigma v \rangle$ in cosmologies with an EMDE that would result from future measurements of $\delta T_{21}$ at $z = 17$.  We also consider how such DM annihilation affects the global signal at other redshifts and the power spectrum of 21\nobreakdash-cm anisotropies.  
We find that measurements of the 21\nobreakdash-cm power spectrum could significantly improve constraints on DM annihilation following an EMDE because the heating of the IGM by DM is more homogeneous than heating by astrophysical sources, which removes the peak in the 21\nobreakdash-cm power spectrum at $z\simeq14$ \cite{Evoli:2014pva, Lopez-Honorez:2016sur, Sun:2023acy, Facchinetti:2023slb}.

This paper is structured as follows. In Section~\ref{section:howEMDE} we discuss how an EMDE enhances the DM annihilation rate and the resulting impact on the IGM temperature. In Section~\ref{section:global}, we use possible measurements of the global 21\nobreakdash-cm signal at $z=17$  to forecast constraints on the DM annihilation cross section following an EMDE. We demonstrate how such scenarios alter the redshift evolution of the global signal and the 21\nobreakdash-cm power spectrum in Sections~\ref{section:Zeus_global} and \ref{section:Zeus_ps}. In Section~\ref{section:conclusion}, we present our conclusions. We use natural units with $\hbar = c = 1$, and we adopt {\it Planck} 2018 cosmological parameters: $h = 0.6766$, $\Omega_m = 0.3111$, and $n_s = 0.9665$~\cite{Planck2018_VI}.

\section{How an EMDE affects the IGM temperature}\label{section:howEMDE}

\subsection{The EMDE Power Spectrum}\label{section:Pk}

 An EMDE sourced by an unstable heavy particle is a generic prediction of numerous early universe scenarios~\cite{Allahverdi2021}. As the energy density of radiation dilutes more rapidly than that of nonrelativistic particles, any sufficiently long-lived heavy particle could eventually dominate the energy density of the universe. This particle, which we will denote as $\phi$, must decay into SM particles prior to neutrino decoupling, at which point the EMDE transitions to a radiation-dominated era (RDE). It is useful to define the reheat temperature $T_\text{RH}$ as the temperature of the universe when the decay rate of $\phi$, $\Gamma_\phi$, is equal to the Hubble rate during radiation domination:
\begin{equation}
	\Gamma_\phi \equiv \sqrt{\frac{8 \pi^3 g_{*,\text{RH}}}{90}}\frac{T_\text{RH}^2}{m_\text{Pl}},
\end{equation}
where $g_{*\text{RH}}$ is the number of relativistic degrees of freedom evaluated at $T_\text{RH}$ and $m_\text{Pl}$ is the Planck mass. Light-element abundances~\cite{Kawasaki1999,Hannestad2004,Ichikawa2005} and the neutrino density measured from the CMB~\cite{Ichikawa2007,DeBernardis2008} require that \mbox{$T_\text{RH} > 8$~MeV}~\cite{deSalas2015,Hasegawa2019}.\footnote{References~\cite{deSalas2015} and \cite{Hasegawa2019} define a reheat temperature $T_\text{RH,BBN}$ such that $\Gamma_\phi = 3 H_\text{RD}(T_\text{RH,BBN})$, where $H_\text{RD}$ is the Hubble rate during a RDE. Hence, the $T_\text{RH,BBN}>4.7$~MeV bound from Ref.~\cite{deSalas2015} translates to $T_\text{RH}>8.1$~MeV.}

During the EMDE, subhorizon DM density perturbations $\delta \equiv \delta \rho_\chi / \langle \rho_\chi \rangle$ grow linearly with the scale factor $a$, while during radiation domination, these perturbations grow logarithmically with $a$. Consequently, DM density perturbations on scales that enter the horizon during the EMDE are enhanced \cite{Erickcek2011,Barenboim2013,Fan2014}. The wavenumber of the mode that enters the horizon at $T_\text{RH}$ is approximately~\cite{Erickcek2015}
\begin{equation} 
    k_\text{RH} \equiv 0.0117
    \left( \frac{T_\text{RH}}{1\text{ MeV}}\right)
    \!\left(\frac{10.75}{g_{*S,\text{RH}}}\right)^{\frac{1}{3}}
    \!\left(\frac{g_{*\text{RH}}}{10.75} \right)^{\frac{1}{2}}
    \text{ pc}^{-1},
\end{equation}
where $g_{*S,\text{RH}} \equiv s/[(2\pi^2/45)T_\text{RH}^3]$ is the effective number of degrees of freedom that contribute to the entropy density $s$ evaluated at $T_\text{RH}$. The linear growth of subhorizon perturbations during an EMDE
leads to a $\mathcal{P}(k) \propto k^{(n_s+3)}$ rise in the dimensionless matter power spectrum for $k>k_\text{RH}$. 
The $k^{(n_s+3)}$ scaling in the power spectrum continues until $k$ exceeds a small-scale cutoff scale $k_\text{cut}$.  We define $k_\text{cut}$ such that the power spectrum is proportional to $\exp[-(k/k_\text{cut})^2]$. If DM interacts with the SM~\cite{Erickcek2011,Erickcek2015,Erickcek2016,Waldstein2017} or is generated nonthermally~\cite{Fan2014,Miller2019}, this small-scale cutoff is usually determined by the DM free-streaming length. If the DM particles are sufficiently cold, the pressure of the $\phi$ particles that dominate during the EMDE, arising from either their standard thermal motion~\cite{Ganjoo2023} or cannibalistic number-changing self-interactions~\cite{Erickcek2021,Erickcek2022}, sets $k_\text{cut}$.  If the EMDE persists long enough for density perturbations to approach the nonlinear regime, the gravitational acceleration of the dark matter particles effectively increases their temperature and inhibits structure formation after the EMDE \cite{Blanco2019, Barenboim:2021swl, Ganjoo:2023fgg}.  This gravitational heating leads a new cutoff scale that can exceed the original cutoff scale in the linear power spectrum by several orders of magnitude \cite{Ganjoo:2023fgg}.

\begin{figure}
\centering
\includegraphics[width=0.45\textwidth]{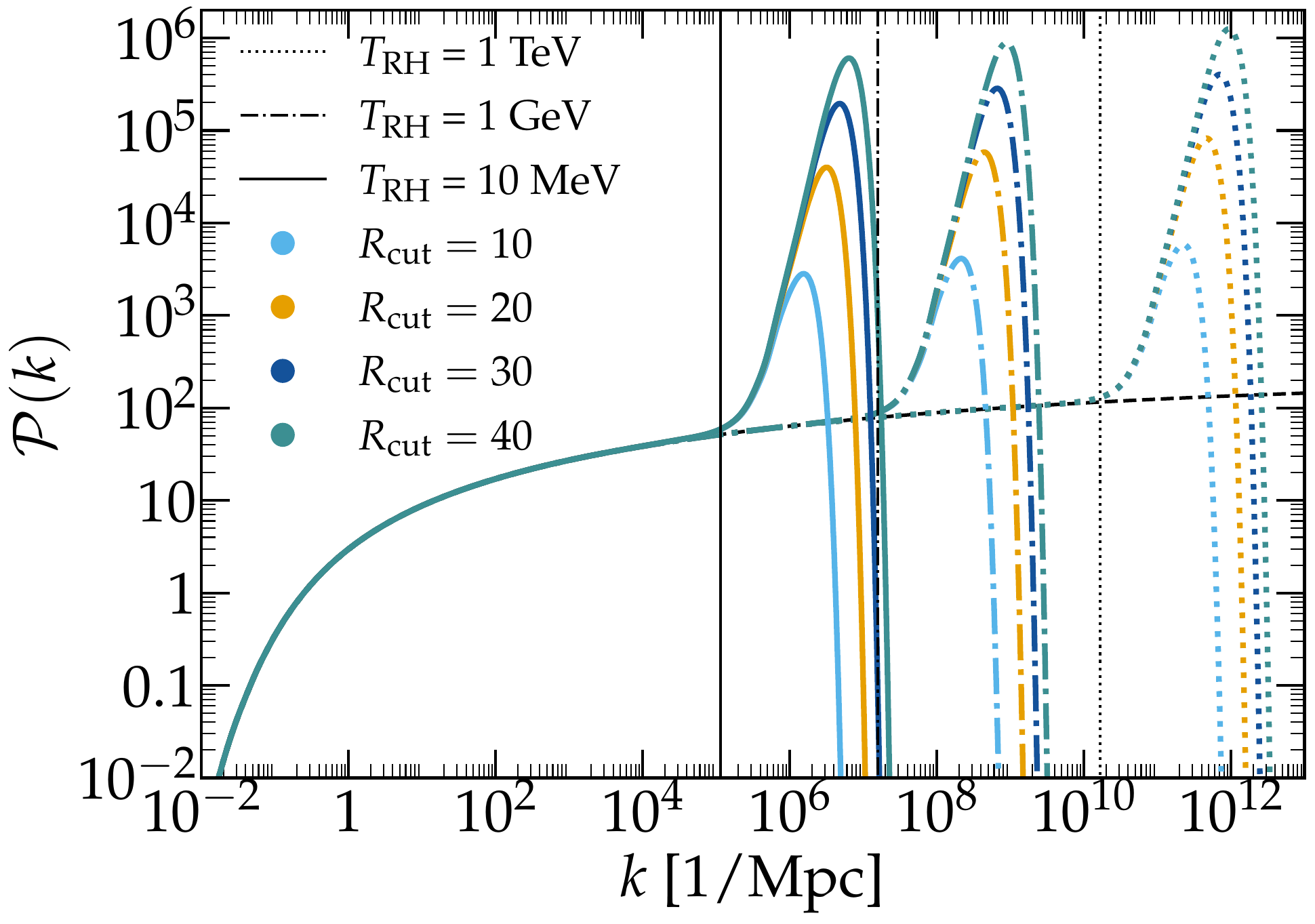}
\caption{
	The dimensionless power spectrum $\mathcal{P}(k)$ of DM density perturbations $\delta(k,a)/a^{0.901}$ during the standard matter-dominated era following EMDEs with various $T_\text{RH}$ and $R_\text{cut} \equiv k_{\text{cut}}/k_{\text{RH}}$. The dashed line shows $\mathcal{P}(k)$ without an EMDE. Vertical lines indicate the corresponding $k_\text{RH}$ for each reheat temperature. Modes with $k>k_\text{RH}$ enter the horizon during an EMDE, while modes with $k>k_\text{cut}$ are suppressed due to the small-scale cutoff to $\mathcal{P}(k)$.}
\label{fig:pk}
\end{figure}

Since baryons do not accrete onto structures at scales $k \gtrsim 10^3$~Mpc$^{-1}$~\cite{Bertschinger2006}, DM density perturbations on these scales grow as $a^{0.901}$ during the standard matter-dominated era~\cite{Hu1996,Delos2019_DarkDegen}. Figure~\ref{fig:pk} shows the dimensionless power spectra of $\delta(a)/a^{0.901}$ evaluated during matter domination for various EMDE scenarios. We apply transfer functions from Ref.~\cite{Erickcek2015} to the matter power spectrum at $z = 500$ generated using the Boltzmann solver \verb|CLASS|~\cite{Blas2011} to produce these power spectra. Figure~\ref{fig:pk} shows power spectra for three reheat temperatures: in each spectrum, there is a bump on scales between $k_\text{RH}$ and $k_\text{cut}$.  Figure~\ref{fig:pk} demonstrates how the ratio $R_\mathrm{cut} \equiv k_\text{cut}/k_\text{RH}$ sets the maximum amplitude of the matter power spectrum.  Figure~\ref{fig:pk} also shows that the maximum amplitude of $\mathcal{P}(k)$ has a logarithmic dependence on $T_\text{RH}$: increasing $T_\text{RH}$ extends the duration of the post-EMDE radiation-dominated era during which the enhanced perturbation modes grow logarithmically with the scale factor. 

Due to the enhancement of small-scale perturbations, peaks in the density field collapse and form gravitationally bound structures earlier after an EMDE than in standard cosmology~\cite{Erickcek2011,Erickcek2015}. In the spherical collapse model, the scale factor at the time of peak collapse $a_\text{sc}$ is set by $\delta(a_\text{sc}) = \delta_c = 1.686$. We adopt the ellipsoidal collapse model and assume that peaks in the density field collapse when the scale factor is equal to $a_\text{ec}$, where $f_\text{ec} \equiv a_{ec}/a_{sc}$ depends on the ellipticity $e$ and prolateness $p$ of the gravitational potential near the precursor overdensity~\cite{Sheth1999}:
\begin{equation}
	f_\text{ec} (e,p) = 1+0.47 [5(e^2-p|p|)f_\text{ec}^2(e,p)]^{0.615}.
	\label{eq:fec}
\end{equation}
However, Eq.~(\ref{eq:fec}) has no solution when $e^2 - p |p| > 0.26$. Peaks with such elliptical tidal fields generally have low heights, and their contributions to the boost factor are negligible~\cite{Delos2022_cuspAnnihilationEffect}. Hence, we assume that peaks without a solution to Eq.~(\ref{eq:fec}) do not collapse.

\begin{figure}
\centering
\subfloat{\includegraphics[width=0.45\textwidth]{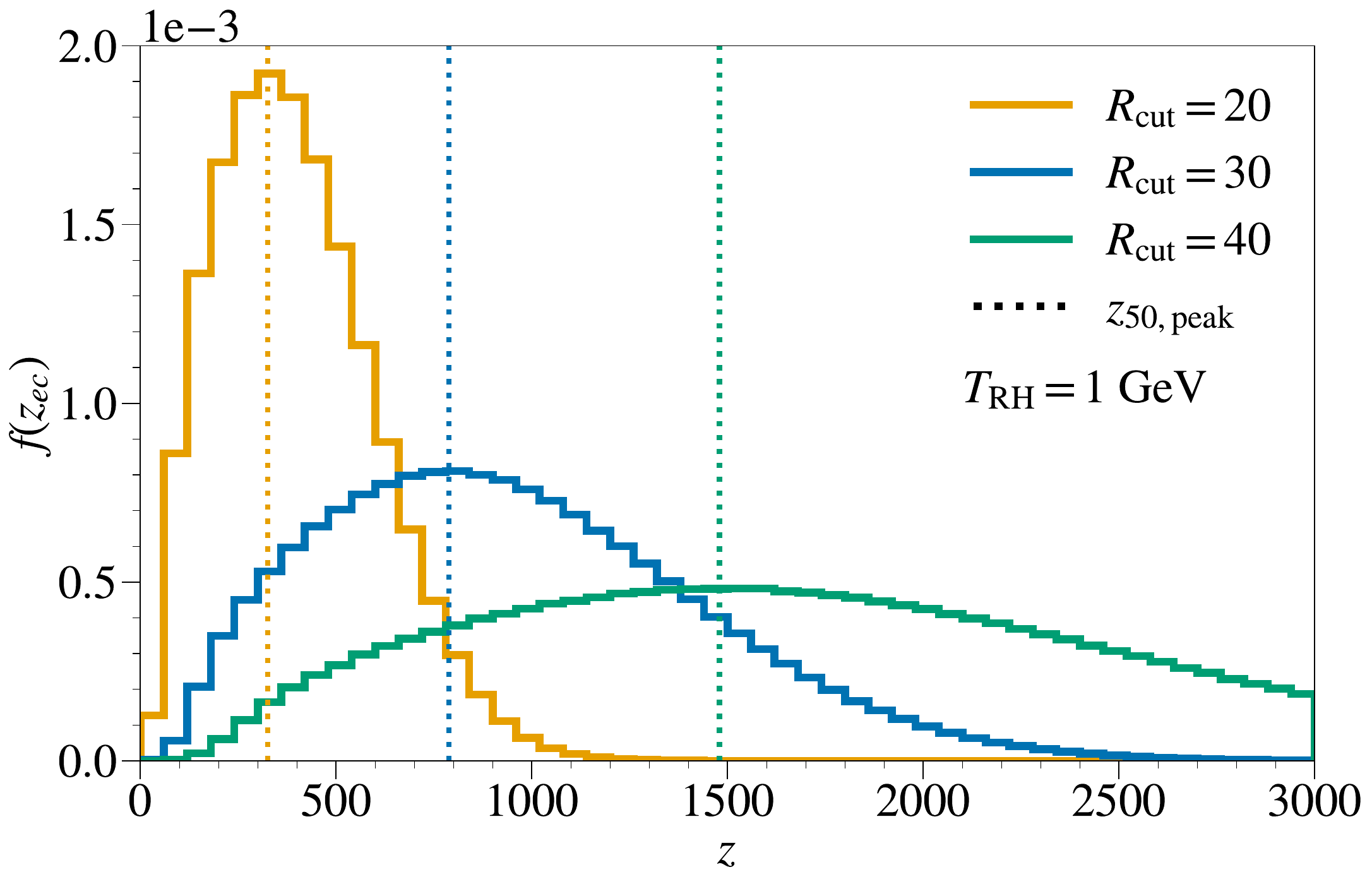}}\\
\subfloat{\includegraphics[width=0.45\textwidth]{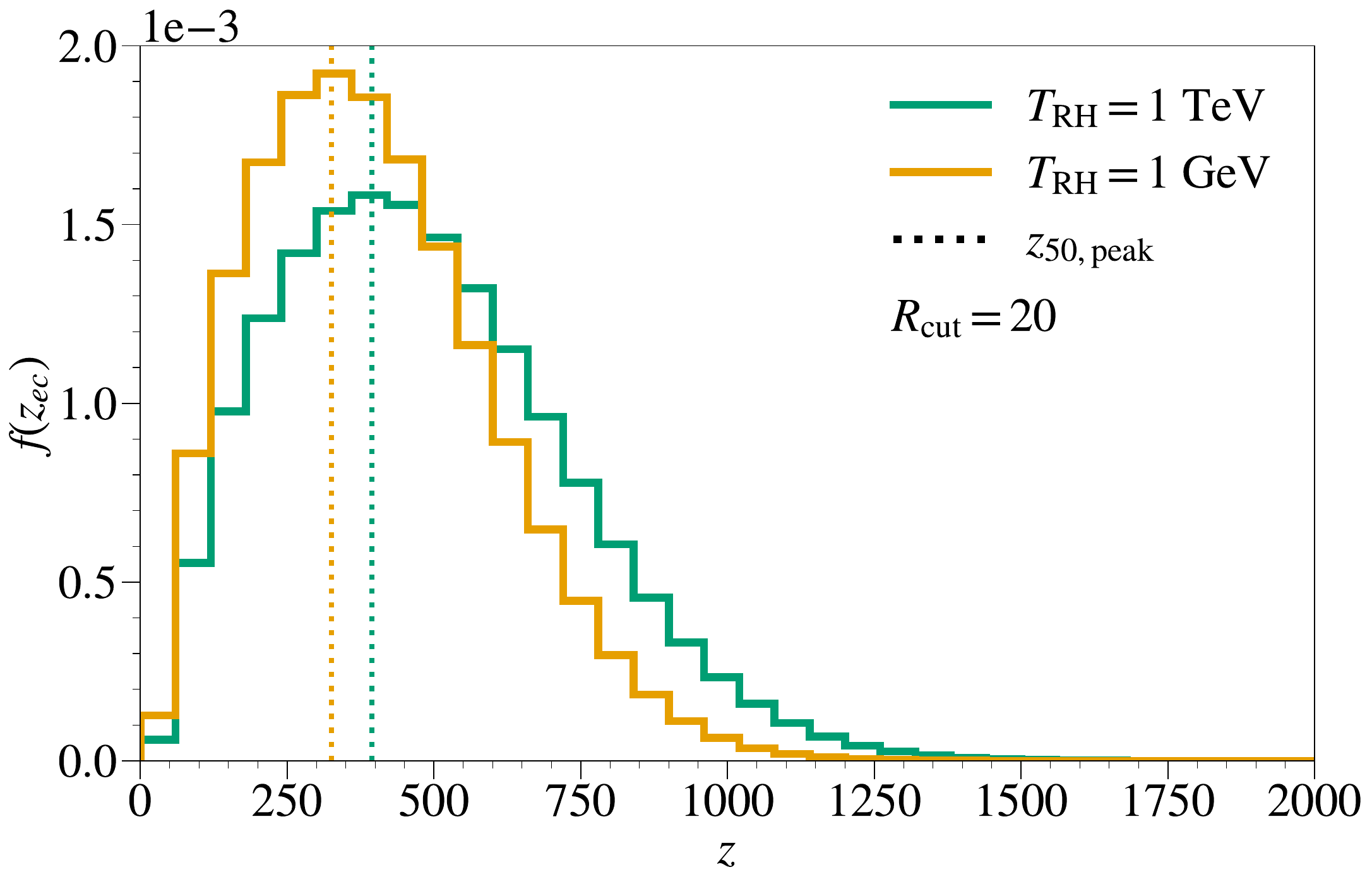}}
\caption{Probability density of collapse redshifts $z_{ec}$ of peaks in the initial density field, calculated from a sample of $N=10^8$ peaks. Dotted lines mark the redshift at which $50\%$ of peaks have collapsed. For an EMDE cosmology with $R_\text{cut} = 10$ and $T_\text{RH} = 1$~GeV, $z_\text{50,peak} = 72$, so it is not displayed in the top panel.}
\label{fig:zcoll}
\end{figure}

Figure~\ref{fig:zcoll} illustrates the distribution of \mbox{$z_\text{ec} = 1/(a_\text{ec})-1$} for a sample of $N = 10^8$ peaks in the initial density field, as well as redshifts $z_\text{50,peak}$ at which half of the peaks in the initial density field have collapsed to form microhalos. For a reheat temperature of 1~GeV, $z_\text{50,peak} = 72$  for $k_{\text{cut}}/k_{\text{RH}} = 10$, while $z_\text{50,peak} =326$ for $k_{\text{cut}}/k_{\text{RH}} = 20$. Microhalos begin forming even earlier for higher cut ratios, with $z_\text{50,peak} =788$ for $k_{\text{cut}}/k_{\text{RH}} = 30$, and \mbox{$z_\text{50,peak} =1479$} for $k_{\text{cut}}/k_{\text{RH}} = 40$. Meanwhile, $z_\text{50,peak}$ is less sensitive to $T_\text{RH}$, as expected from the weaker influence of $T_\text{RH}$ on the maximum of $\mathcal{P}(k)$. Increasing the reheat temperature to $T_\text{RH} = 1$~TeV leads to $z_\text{50,peak} =86$ for $k_{\text{cut}}/k_{\text{RH}} = 10$ and $z_\text{50,peak} =1792$ for $k_{\text{cut}}/k_{\text{RH}} = 40$. These early-forming microhalos have high central densities, resulting in a significant impact on the DM annihilation rate.

\subsection{Calculating the Boost Factor}\label{section:Boost}

To determine the effect of EMDE-enhanced structure formation on the DM annihilation rate, we calculate the boost factor due to halos
\begin{equation}\label{eq:Boost}
    \mathcal{B}(z) \equiv 
    \frac{\langle \rho_\chi^2(z) \rangle_h}
    {\langle \rho_\chi(z) \rangle^2 },
\end{equation}
where $\langle \rho_\chi \rangle$ is the average DM density, and $\langle \rho_\chi^2(z) \rangle_h$ is the average of the DM density squared within halos. If $n_\text{halo}(z)$ is the comoving number density of halos at $z$, 
\begin{equation} \label{eq:rho2_nhalo}
    \langle \rho_\chi^2(z) \rangle_h = n_\text{halo}(z) (1+z)^3  \langle J(z) \rangle,
\end{equation}
where
\begin{equation}\label{eqJFactor}
    J \equiv \int \rho^2 dV
\end{equation}
is the $J$ factor for each microhalo, and $ \langle J(z) \rangle$ is the average $J$ factor of halos that are present at a redshift $z$.

We use the Peak-to-Halo (P2H) method~\cite{Delos2019_PredictDens} to predict the properties of the first DM halos and to calculate the resulting boost to DM annihilation.
First-generation halos in n-body simulations have $\rho\propto r^{-3/2}$ inner density profiles~\cite{Ishiyama2010,Ishiyama2014,Anderhalden2013,Polisensky:2015,Ogiya:2016hyo,Angulo2017,Delos2018_Are,Delos2018_Density,Ishiyama:2019hmh,Colombi:2020xbv}, and these ``prompt cusps'' form out of the collapse of peaks in the initial density field with properties tightly linked to those of the initial peaks~\cite{Delos2019_PredictDens,Delos2022_cuspSurvival,Ondaro-Mallea:2023qat}. In the P2H method, we use Gaussian statistics~\cite{Bardeen1986} to obtain the distribution of initial density peaks, and we associate each peak that reaches the ellipsoidal collapse threshold of Ref.~\cite{Sheth1999} with a collapsed, gravitationally bound halo with inner density profile given by $\rho = A r^{-3/2}$. The coefficient $A$ is determined by the local properties of the initial density peak, namely its amplitude $\delta$ and its curvature $\nabla^2\delta$~\cite{Delos2019_PredictDens}. 
For halos that form after matter-radiation equality and are smaller than the baryonic Jeans mass,
$A$ can be predicted in terms of $\tilde\delta(\vec x)\equiv\delta(\vec x,a)/a^{0.901}$ \cite{Delos2019_DarkDegen}:
\begin{equation}
\label{eq:A}
A = \alpha \delta_c^{\frac{3}{2}(1-\frac{1}{\mu})}\langle \rho_{\chi,0}\rangle\tilde{\delta}^{\frac{3}{4}(\frac{2}{\mu}+1)}|\nabla^2\tilde{\delta}|^{-3/4} f_\text{ec}^{-\frac{3}{2\mu}}(e,p),
\end{equation}
where $\alpha = 12.1$, $\mu = 0.901$, and $\langle \rho_{\chi,0}\rangle$ is the current average density of DM.

Recent simulations indicate that all halos retain these $\rho=Ar^{-3/2}$ 
prompt cusps in their centers \cite{Delos2022_cuspSurvival,Delos:2025pen}. Each cusp extends out to a radius $r_\mathrm{cusp}$, which is set by the physical size of the initial collapsing region, and down to a minimum radius $r_\mathrm{core}$ set by the initial DM phase-space density, below which it must shallow into a finite-density core due to Liouville's theorem.\footnote{Dark matter annihilation would separately impose a maximum density $\rho_\mathrm{max}\sim m_\chi/(\langle \sigma v \rangle t)$, where $t$ is the age of the Universe. Typically, however, this $\rho_\mathrm{max}$ exceeds the core density from Liouville's theorem by many orders of magnitude \cite{Ganjoo:2024hpn}.} If the transition from cusp to central core is abrupt, then
\begin{equation}\label{eq:JFactorNFW}
   J \simeq (4\pi/3)\omega A^2,
\end{equation}
where $\omega=1+3\log(r_\mathrm{cusp}/r_\mathrm{core})$. Standard cosmologies with WIMP DM lead to $r_\mathrm{cusp}/r_\mathrm{core}\approx 500$, corresponding to $\omega\approx 19$ \cite{Delos2022_cuspAnnihilationEffect}.
Since altering the DM temperature or mass tends to affect $r_\mathrm{cusp}$ (through the free-streaming length) and $r_\mathrm{core}$ (through the phase-space density) in concert, $r_\mathrm{cusp}/r_\mathrm{core}$ is not very sensitive to DM properties. However, the addition of an EMDE can alter $r_\mathrm{cusp}/r_\mathrm{core}$ more significantly. For example, the pressure of the dominant $\phi$ particle can set $k_\mathrm{cut}$ (and hence $r_\mathrm{cusp}$) independently from the DM microphysics that sets the phase-space density (and hence $r_\mathrm{core}$), potentially leading to much larger $r_\mathrm{cusp}/r_\mathrm{core}$. More broadly, the modified expansion history during an EMDE leads to larger $r_\mathrm{cusp}/r_\mathrm{core}$, while the earlier halo formation that results from an EMDE leads to smaller $r_\mathrm{cusp}/r_\mathrm{core}$. Generally, since $\omega$ is logarithmic in $r_\mathrm{cusp}/r_\mathrm{core}$, $\omega$ is expected to be of order 10, but its precise value depends on the details of the cosmological scenario~\cite{Ganjoo:2024hpn}. 
Additionally, Refs.~\cite{Delos2022_cuspAnnihilationEffect,Ganjoo:2024hpn} found that about half of the initial peaks that are predicted to collapse can be associated with surviving cusps at late times. The other peaks either fail to collapse because they accrete onto other halos first, or they form prompt cusps that are subsequently lost in halo mergers.

Since our primary objective is to evaluate how well 21\nobreakdash-cm observations can probe EMDE scenarios, we wish to compare our results to the constraints presented in Ref.~\cite{Delos2019_DarkDegen}, which used observations of the isotropic gamma-ray background (IGRB) to constrain the same post-EMDE power spectra, specified by $T_\mathrm{RH}$ and $k_\mathrm{cut}$, that we consider. 
Therefore, we apply the P2H method under the extremely conservative assumptions of Ref.~\cite{Delos2019_DarkDegen}. Contrary to more recent results~\cite{Delos2022_cuspSurvival},
Ref.~\cite{Delos2019_DarkDegen} assumed that repeated halo mergers drive the $\rho \propto r^{-3/2}$ profile toward the shallower Navarro-Frenk-White (NFW) density profile, \mbox{$\rho(r) = \rho_s / [(r/r_s)(1+r/r_s)^2]$}~\cite{Navarro1995,Navarro1996}. If the transition from the $\rho \propto r^{-3/2}$ profile to the NFW profile conserves mass within the scale radius $r_s$, the $J$ factor is given by Eq.~(\ref{eq:JFactorNFW}) with $\omega = 1.8$, and this was the $J$ factor used in Ref.~\cite{Delos2019_DarkDegen}.
Ref.~\cite{Delos2019_DarkDegen} also calculated the boost factor using the full complement of peaks, based on simulation results that suggest that mergers of NFW halos approximately preserve the total annihilation luminosity \cite{Drakos2019,Delos2019_PredictDens}.
To facilitate comparisons with constraints from the IGRB, we also take $\omega = 1.8$ and include every peak after its predicted collapse time. However, based on the discussion above, we are likely underestimating the annihilation luminosity of each halo by a factor of about 10 (since $\omega$ should be closer to 19) and hence the overall boost factor by a factor of about 5 (accounting for the fraction of peaks that can be associated with surviving objects).

To calculate the comoving number density of halos in Eq.~(\ref{eq:rho2_nhalo}), we must obtain the comoving number density of peaks in the density field. The number density of peaks can be computed from the spectral moments of the power spectrum~\cite{Bardeen1986}:
\begin{align}
    \sigma_j &\equiv \left( \int_{0}^{\infty}\frac{\mathrm{d}k}{k}\mathcal{P}(k)k^{2j}\right )^{1/2}.
\end{align}
The peak distribution in $\nu \equiv \tilde{\delta} / \sigma_0$ and \mbox{$x \equiv -\nabla ^2 \tilde{\delta}/\sigma_2 $} is
\begin{equation}
\label{eq:dndnudx}
\frac{\mathrm{d}^2n}{\mathrm{d} \nu \mathrm{d}x} = \frac{\mathrm{e}^{-\nu^2/2}}{(2\pi)^2 R_*^3}f(x) \frac{\exp[-\frac{1}{2}(x-\gamma\nu)^2/(1-\gamma^2)]}{[2\pi(1-\gamma^2)]^{1/2}},
\end{equation}
where $R_*~\equiv~\sqrt{3} \sigma_1/\sigma_2$, $\gamma~\equiv~\sigma_1^2/(\sigma_0 \sigma_2)$, and
\begin{align}
f(x) \equiv &
    \sqrt{\frac{2}{5\pi}}
    \left[ \left(\frac{31}{4}x^2+\frac{8}{5}\right)\mathrm{e}^{-\frac{5}{8}x^2}+\left(\frac{x^2}{2}-\frac{8}{5} \right)\mathrm{e}^{-\frac{5}{2}x^2}\right]  \nonumber \\ 
    &+\frac{x^3-3x}{2}\left[ \mathrm{erf}\left(\sqrt{5/2}x\right)+\mathrm{erf}\left(\sqrt{5/8}x\right) \right].
\end{align}
Integrating Eq.~(\ref{eq:dndnudx}) over $\nu\geq0$ and $x\geq0$ yields the total comoving number density of peaks $n$.

We use the distribution of $\nu, x, e,$ and $p$ for peaks in Gaussian random fields to generate a sample of $N = 10^8$ density peaks from our EMDE power spectra following Refs.~\cite{Delos2019_PredictDens,Bardeen1986}. Parameters $\nu$ and $x$ are simultaneously sampled from Eq.~(\ref{eq:dndnudx}) using rejection methods. Once this sample is obtained, $e$ and $p$ can then be sampled from the conditional cumulative distributions provided in Ref.~\cite{Sheth1999} by inverse transform sampling. Finally, these peak parameters are used to calculate $A$ for each halo using Eqs.~(\ref{eq:fec}) and~(\ref{eq:A}). 

For a sample of $N$ density peaks from a primordial field with a comoving number density of peaks $n$, Eq.~(\ref{eq:rho2_nhalo}) can be written as
\begin{equation} \label{eq:rho2}
    \langle \rho_\chi^2(z) \rangle_h = \frac{n(1+z)^3}{N} \sum_{z_{f,i}>z} J_i, 
\end{equation}
where $z_{f,i}$ is the redshift at which the $i$th halo formed, and $J_i$ is its $J$ factor. Results from n-body simulations suggest that, 
for NFW halos, the sum of the $J$ factors does not vary much as halos
merge~\cite{Delos2019_DarkDegen,Drakos2019}. Taking advantage of this consistency, we evaluate $\Sigma J$ based on the P2H-predicted halo population, with $J_i = (4\pi/3)\omega A_i^2$ for each density peak. N-body simulations also indicate that the inner density profiles of the halos that form from the direct collapse of overdense regions closely match their final inner density profiles when  $a=1.15a_{\text{ec}}$~\cite{Delos2019_PredictDens}. We therefore assign a halo formation redshift $z_f \equiv 1/(1.15a_{\text{ec}})-1$ to peaks that collapse at $a_{\text{ec}}$. We expect proto-halos to boost the annihilation rate at redshifts higher than $z_f$, so this choice of $z_f$ leads to a conservative value of $\mathcal{B}(z)$ before the halo population is established.

\begin{figure}
\includegraphics[width=0.48\textwidth]{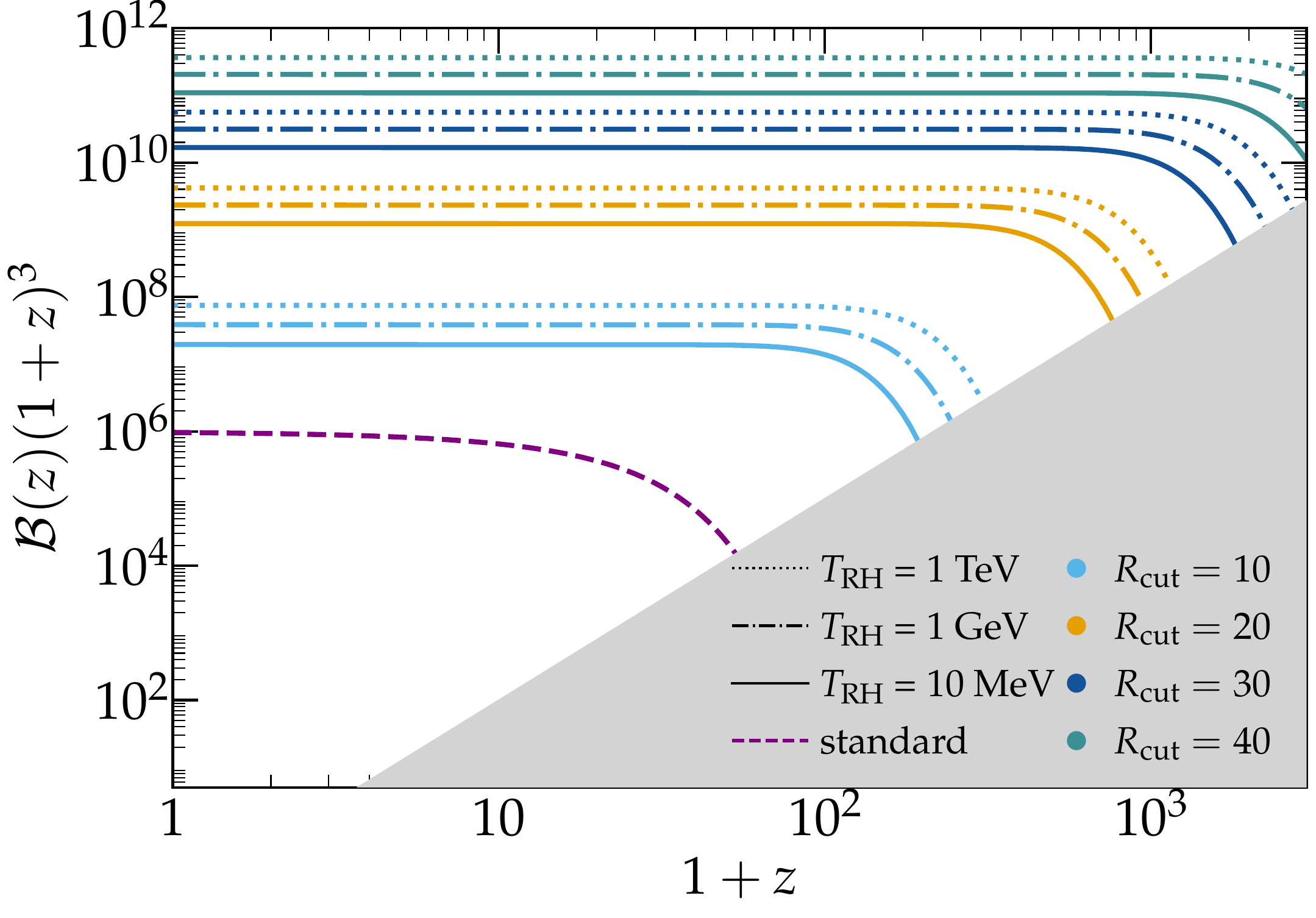}
\caption{The boost factor $\mathcal{B}(z)(1+z)^3$ as a function of redshift for various EMDE cosmologies. The shaded region indicates where $\mathcal{B}(z) \leq 0.1$. The standard boost shows Eq.~(\ref{eq:erf}) with parameters $z_h = 30$, \mbox{$b_h = 10^6$}, and $\beta = 3$.}
\label{fig:boost}
\end{figure}

As density peaks reach their collapse thresholds, the number of halos that contribute to the sum of $J$ factors in Eq.~(\ref{eq:rho2}) increases. The inclusion of later-forming halos, however, does not significantly alter this sum because earlier-forming halos have larger $A$ values. As a result, this sum stops growing at some point, after which the boost factor \mbox{$\mathcal{B}(z) \propto (1+z)^{-3}$} as seen from Eqs.~(\ref{eq:Boost}) and~(\ref{eq:rho2}).  It is therefore useful to consider the evolution of \mbox{$\mathcal{B}(z)(1+z)^3$}, which is shown in Fig.~\ref{fig:boost}. For each EMDE scenario, the plateau in $\mathcal{B}(z)(1+z)^3$ reflects the establishment of the halo population that dominates the annihilation rate. Recalling the $z_\mathrm{peak}$ values shown in Fig.~\ref{fig:zcoll}, we see that \mbox{$\mathcal{B}(z)(1+z)^3$} is nearly constant for $z\lesssim z_\mathrm{peak}$. Therefore, the halos that form after $z_\mathrm{peak}$ do not significantly increase the boost factor.

As expected by the small effect of $T_\text{RH}$ on the maximum of $\mathcal{P}(k)$, $\mathcal{B}(z)(1+z)^3$ depends weakly on the reheat temperature. In contrast, the peak amplitude of the power spectra, which sets the formation time of the first halos, is highly sensitive to the ratio $k_{\text{cut}}/k_{\text{RH}}$, as shown in Fig.~\ref{fig:zcoll}. We also see this sensitivity to $k_{\text{cut}}/k_{\text{RH}}$ in the boost factors. Since most halos form at high $z$ for high $k_{\text{cut}}/k_{\text{RH}}$, $\mathcal{B}(z)(1+z)^3$ reaches its maximum value as early as $z\approx 2000$ for $k_{\text{cut}}/k_{\text{RH}} = 40$. For $k_\text{cut}/k_\text{RH} = 40$, $\mathcal{B} (z = 1000) > 100$ for all reheat temperatures above 10~MeV.

In contrast, no boost to annihilation from halos is expected at $z\gtrsim 60$ in standard structure formation scenarios, and the value of the boost factor today is not expected to exceed $\sim\!10^6$ \cite{Poulin2015,Fermi-LAT:2015qzw, Delos2022_cuspAnnihilationEffect}. 
For standard scenarios of structure formation, the boost factor can be computed using Press-Schechter halo mass functions ~\cite{Press1974} with the assumption that all halos have NFW profiles (without $\rho\propto r^{-3/2}$ cusps):
\begin{equation}
\label{eq:erf}
    \mathcal{B}(z) = \frac{b_h}{(1+z)^{\beta}} \text{erfc} \left( \frac{1+z}{1+z_h} \right),
\end{equation}
where $\beta, z_h$, and $b_h$ are free parameters~\cite{Evoli2014}. The characteristic redshift $z_h$ is the redshift at which the first halos that formed at some $z_f\approx 2z_h$ start to contribute to the boost, and $b_h$ is slightly greater than the value of the boost today. We plot the dashed line in Fig.~\ref{fig:boost} using $z_h = 30, b_h = 10^6, \beta = 3$ to show the maximal boost factor expected in the absence of an EMDE \cite{Poulin2015}.

The P2H prediction for the halo inner density profile is only valid for halos that form during matter domination. Thus, we restrict the cut ratio to $k_{\text{cut}}/k_{\text{RH}}~\leq~40$ in our work and make the approximation that the universe is matter dominated for $z \leq 3000$. For $k_{\text{cut}}/k_{\text{RH}}~=~40$ and $T_\text{RH} = 1$~GeV, $\mathcal{B}(z = 3000)$ is less than $1\%$ of the boost today, and only $8.5\%$ of peaks have collapsed by this time. Higher cut ratios are possible~\cite{Blanco2019,Erickcek2016,Ganjoo2023}, and they further enhance density fluctuations. These fluctuations lead to locally matter-dominated regions during the RDE, in which halos can form prior to matter-radiation equality~\cite{Blanco2019,StenDelos:2022jld,Delos:2023fpm}. However, the density profiles of these structures are unknown. Further work is needed to compute the boost due to microhalos that form during radiation domination, but we expect that constraints on $k_\text{cut}/k_\text{RH} = 40$ apply to all larger cut ratios. 


\subsection{Energy Injection}\label{section:EInj}

Primary particles released by annihilating DM rapidly cascade into photons, electrons, and positrons. The rate of energy injected per volume by annihilating DM with a velocity-averaged cross section $\langle \sigma v \rangle$ is
\begin{equation}\label{eq:EinjWithBoost}
    \left( \frac{dE}{dVdt} \right)^{\text{inj}} = \langle \rho_{\chi,0} \rangle^2 (1+z)^6 \frac{\langle \sigma v \rangle}{m_\chi} \left[ \mathcal{B}(z) + f_u \right],
\end{equation}
where $m_\chi$ is the DM particle mass, and $f_u$ is the fraction of DM that is not bound in halos. Before halos form, $\mathcal{B}(z) = 0 $, $f_u = 1$, and the energy injected from the DM in the homogeneous background dominates Eq.~(\ref{eq:EinjWithBoost}). Following an EMDE, the fraction of free DM particles approaches zero once microhalos begin to form \cite{Erickcek2011}. However, since $\mathcal{B}(z) \gg 1$ after the formation of the first halos, we neglect the evolution of the unbound DM fraction and take $f_u = 1$ in Eq.~(\ref{eq:EinjWithBoost}). 

Recalling that \mbox{$\mathcal{B}(z) \propto (1+z)^{-3}$} after about half of the density peaks have collapsed, Eq.~(\ref{eq:EinjWithBoost}) implies that the energy injection rate after an EMDE is proportional to $(1+z)^3$ for $z\lesssim z_\mathrm{peak}$. This is the same redshift dependence as the energy injection rate for decaying dark matter, and the correspondence between DM decay and annihilation within microhalos has been used to convert IGRB bounds on decaying dark matter into bounds on DM annihilation following an EMDE~\cite{Blanco2019,Delos2019_DarkDegen,Ganjoo:2024hpn}. If the 21\nobreakdash-cm signatures of annihilating DM only depend on energy injected when $z\lesssim z_\mathrm{peak}$, the same technique can be applied to 21\nobreakdash-cm constraints on decaying DM, as we discuss in Section~\ref{section:global}.

The injected energy can ionize, excite, or heat the IGM gas. In general, the injected energy is not immediately deposited into the gas, and secondary photons can significantly redshift before energy deposition occurs~\cite{Slatyer2013}. This delay in energy deposition is parametrized by the ``efficiency" functions $f_c$, where $c =$~(heating, ionization, and excitation) is the channel that the energy is deposited to. For each $c$,
\begin{equation}\label{eq:eDep}
  \left( \frac{dE}{dVdt} \right)^{\text{dep}}_c =  f_c(z) \left( \frac{dE}{dVdt} \right)^{\text{inj}}_{\text{smooth}},
\end{equation}
where $(dE/dVdt)^{\text{dep}}_c$ is the energy deposition rate per volume at redshift $z$, while $(dE/dVdt)^{\text{inj}}_{\text{smooth}}$ is the energy injected from the smooth DM density, i.e., $\mathcal{B}(z) = 0$ in Eq.~(\ref{eq:EinjWithBoost}), at the same redshift. With this definition, the $f_c(z)$ functions account for both the delayed and incomplete absorption of the injected energy and the boost to the annihilation rate due to structure formation. 

The IGM is cooled by adiabatic expansion and inverse Compton scattering from CMB photons~\cite{Peebles1968}:

\begin{equation}
    \dot{T}_K^{(0)} = -2 H T_K + \Gamma_C (T_{\text{CMB}} - T_K), \label{eq:T0}
\end{equation}
where the overdot indicates the proper time derivative, $^{(0)}$~denotes the absence of energy injection, and $H$ is the Hubble rate. The Compton scattering rate is
\begin{equation}
    \Gamma_C = \frac{x_e}{1+\mathcal{F}_\text{He}+x_e} \frac{8 \sigma_T a_r T^4_\text{CMB}}{3m_e},
\end{equation}
where $x_e \equiv n_e/n_\text{H}$ is the fraction of free electrons, $\mathcal{F}_{\text{He}} \equiv n_{\text{He}}/n_{\text{H}}$ is the relative abundance of helium nuclei by number, $\sigma_T$ is the Thomson cross section, $a_r$ is the radiation constant, and $m_e$ is the electron mass. Hence, the IGM temperature depends on the free electron fraction.

The evolution of the free electron fraction is~\cite{AliHaimoud2010,Liu2019}
\begin{equation}
    \dot{x}_{\text{e}}^{(0)} = -\mathcal{C} \left[ n_\text{H} x_e x_\text{HII} \alpha_\text{H}
    - 4 (1-x_\text{HII})\beta_\text{H} \text{e}^{-\frac{E_{21}}{T_\text{CMB}}} \right], \label{eq:x0}
\end{equation}
where $\mathcal{C}$ is the Peebles factor, $x_\text{HII}$ is the ratio of the number density of free protons and the number density of both neutral and ionized hydrogen, $\alpha_H(T_K)$ is the case B recombination coefficient, $\beta_H(T_\text{CMB})$ is the photoionization rate, and $E_{21} = 10.2$~eV is the Ly$\alpha$ transition energy. Case B recombination neglects free electrons recombining directly to ground state hydrogen, as this process emits a photon that almost immediately ionizes nearby neutral hydrogen in optically thick clouds, leading to a constant $x_e$.

Accounting for the heating and ionization of the IGM due to energy injection~\cite{Liu2019},
\begin{align}
    \dot{T}_K &= \dot{T}_K^{(0)} + \frac{2 f_{\text{heat}}}{3(1+\mathcal{F}_{\text{He}}+x_e)n_H} \left( \frac{dE}{dVdt} \right)^{\text{inj}},  \label{eq:TEvolution} \\ 
    \dot{x}_{e} &= \dot{x}_{e}^{(0)} + \left[ \frac{f_{\text{H ion}}}{\mathcal{R}n_H}+\frac{(1-\mathcal{C})f_{\text{exc}}}{0.75 \mathcal{R}n_H} \right ]  \left( \frac{dE}{dVdt} \right)^{\text{inj}},  \label{eq:XEvolution}
\end{align}
where $f_c$ are the efficiency functions for heating, ionization, and excitation of the IGM gas, and $\mathcal{R} = 13.6$ eV.
\begin{figure}[t]
    \includegraphics[width=0.45\textwidth]{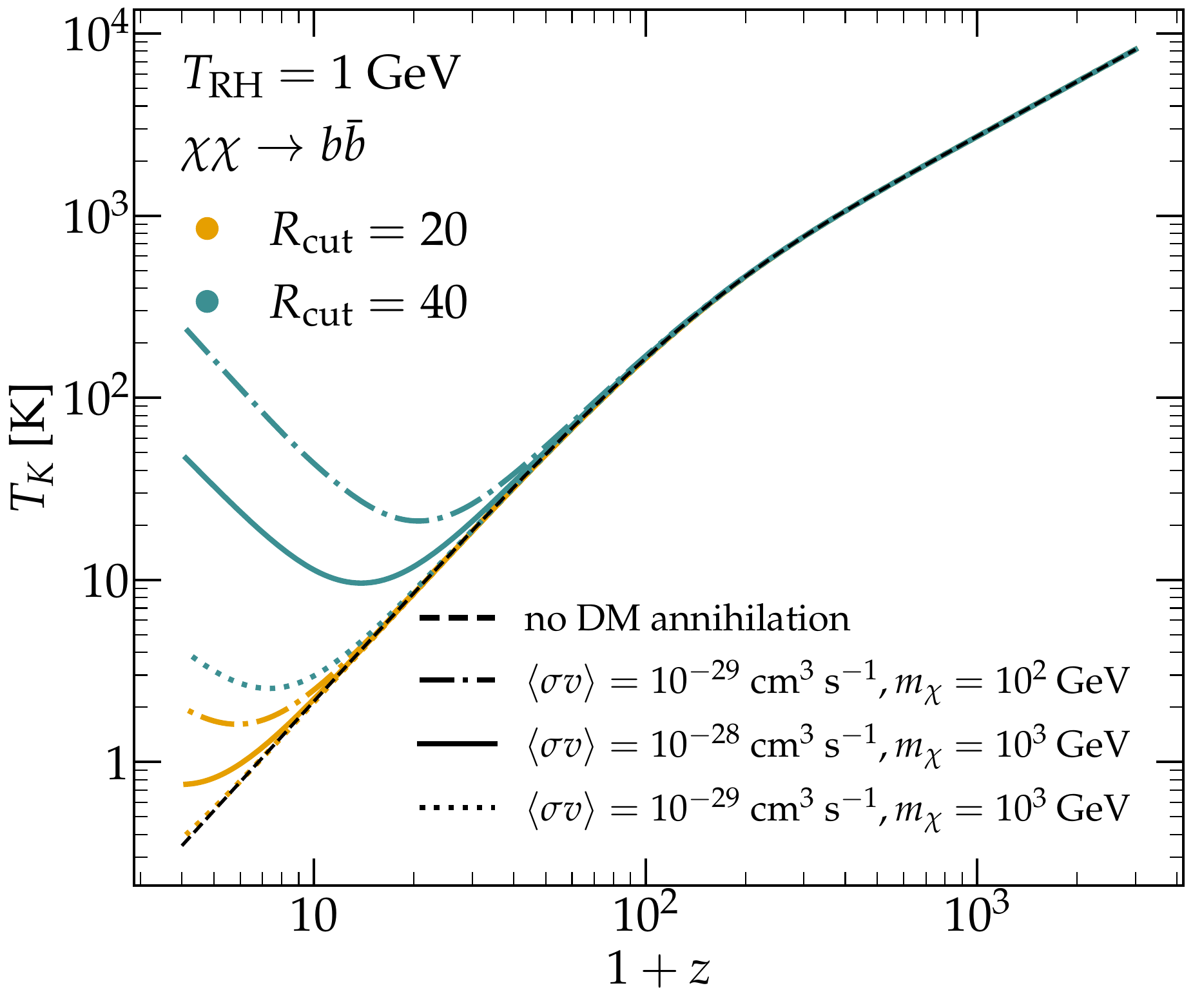}
	\caption{Evolution of the IGM temperature due to heating from DM annihilating into $b\bar{b}$ quarks. DM annihilation within the microhalos that form after an EMDE injects energy into the IGM and increases the IGM temperature after it decouples from the CMB temperature at $z\simeq300$. Energy injection from astrophysical sources is not included here.}
	\label{fig:TEvolution}
\end{figure}
We use the \verb|DarkHistory|~\cite{Liu2019} Python package to solve Eq.~(\ref{eq:TEvolution}) and~(\ref{eq:XEvolution}) including energy injection due to annihilating DM with the boost functions computed in Section~\ref{section:Boost}. Employing the methods developed in Ref.~\cite{Slatyer:2015kla}, \verb|DarkHistory| computes the efficiency functions $f_c$ at every redshift while accounting for changes to the ionization fractions of both hydrogen and helium due to the energy injected by DM.  

Figure~\ref{fig:TEvolution} illustrates the effect of different EMDE cosmologies, DM annihilation cross sections, and DM masses on the IGM temperature. At redshifts above $z\approx300$, the IGM temperature is coupled to the CMB temperature so that $T_K \propto (1+z)$. The boost due to EMDEs can quickly reach a large value for high $k_\text{cut}/k_\text{RH}$; $\mathcal{B}(z = 300)\approx 100$ for $k_\text{cut}/k_\text{RH} = 20$, while $\mathcal{B}(z = 1000)\approx 100$ for $k_\text{cut}/k_\text{RH} = 40$. However, the impact of this large boost is not immediately seen in $T_K$, as inverse Compton scattering efficiently transfers energy to the CMB for $z\gtrsim 300$.
At lower redshifts, $T_K$ adiabatically cools until the energy injected from annihilating DM exceeds the energy lost due to the expansion of the universe.  At this point, $T_K$ begins to increase.  Energy injection from astrophysical sources such as stars and active galactic nuclei are also expected to heat the IGM when $z \lesssim 17$, but contributions from these sources are not shown in Fig.~\ref{fig:TEvolution}.

The rate of energy injection depends on the ratio $\langle \sigma v \rangle/m_\chi$ as seen in Eq.~(\ref{eq:EinjWithBoost}). The dotted curves in Fig.~\ref{fig:TEvolution} have a smaller $\langle \sigma v \rangle/m_\chi$ ratio than the solid and dash-dotted curves, leading to less DM heating of the IGM.  For the same value of $\langle \sigma v \rangle/m_\chi$, DM particles with larger masses are less effective at heating the IGM because it takes longer for higher-energy particles to deposit their energy. This effect is illustrated by the dash-dotted and solid lines in Fig.~\ref{fig:TEvolution}, which have the same value of $\langle \sigma v \rangle/m_\chi$. Figure~\ref{fig:TEvolution} also demonstrates how the height of the peak in the matter power spectrum affects the DM annihilation rate. Increasing $R_\mathrm{cut} \equiv k_\mathrm{cut}/k_\mathrm{RH}$ from 20 to 40 increases the maximum of ${\cal P}(k)$ by a factor of 16, leading to denser and earlier-forming microhalos with higher annihilation rates that heat the IGM at higher redshifts.


\subsection{The 21\nobreakdash-cm Signal}\label{section:21cm}

The global 21\nobreakdash-cm signal depends on the spin temperature $T_S$, which describes the relative populations of the triplet and singlet ground states of neutral hydrogen. Interactions with CMB photons couple $T_S$ to $T_\text{CMB}$, while collisions with other hydrogen atoms, protons, or electrons couple $T_S$ to the kinetic temperature $T_K$ with a coupling coefficient $x_c$. 
Finally, the absorption and re-emission of Ly$\alpha$ photons couple $T_S$ to $T_K$ through the Wouthuysen-Field (WF) effect \cite{Wouthuysen1952, Field1958}, with a coupling coefficient $x_\alpha$~\cite{Hirata:2005mz}. The combined influence of these three interactions on the spin temperature is given by~\cite{Field1959, Hirata:2005mz},
\begin{equation}\label{Eq8} T_S^{-1} = \frac{T_\text{CMB}^{-1}+x_{\alpha} T_{K}^{-1}+x_c T_{K}^{-1}}{1+x_{\alpha}+x_c}. \end{equation}
The observed 21\nobreakdash-cm signal is conventionally expressed as the differential brightness temperature $\delta T_{21}$, relative to the CMB~\cite{Zaldarriaga2003},
\begin{align}
    \delta T_{21} =& \frac{T_S - T_{\text{CMB}}}{1+z}(1-e^{-\tau}), \notag \\
     \approx & \frac{T_S - T_{\text{CMB}}}{1+z}\tau, \notag \\
     \approx & x_{\text{HI}} \left( \frac{\Omega_b h^2}{0.020} \right) \left( \frac{0.15}{\Omega_m h^2} \frac{1+z}{10} \right)^{1/2} \notag \\
    & \times \left( \frac{T_S-T_{\text{CMB}}}{T_S} \right) 23\text{ mK},
    \label{eq:deltaT21Long}
\end{align}
where $\tau$ is the optical depth of neutral IGM at the frequency corresponding to $21(1 + z)$~cm, $x_{\text{HI}}$ is the fraction of neutral hydrogen, $\Omega_m$ and $\Omega_b$ are the  matter and baryon abundances, respectively, and $h$ is the Hubble rate today in units of 100 km s$^{-1}$ Mpc$^{-1}$. 
Note that the 21\nobreakdash-cm signal can only be detected when the spin temperature deviates from the CMB temperature; $T_S>T_{\rm CMB}$ will give rise to emission, whereas $T_S<T_{\rm CMB}$ leads to absorption.

Shortly after recombination, the baryon density is high enough for atomic collisions to keep \mbox{$T_S\simeq T_K$}. Meanwhile, Compton scattering keeps $T_K \simeq T_\mathrm{CMB}$ for $z>200$. As the universe expands, both the Compton scattering rate and the atomic collision rate decrease. The gas decouples from the CMB and cools adiabatically: $T_K \propto (1+z)^2$ for $z\lesssim 200$. Radiative interactions dominate the evolution of $T_S$ after $z\simeq 150$, so that $T_S \simeq T_{\text{CMB}}$. However, when star formation starts around $z\approx 20$, the emitted Ly$\alpha$ photons cause $T_S$ to approach $T_K$ through the WF effect. 
When X-ray sources begin to heat the IGM at $z\approx15$~\cite{Furlanetto2006,Cohen2017}, both $T_K$ and $T_S$ begin to increase and eventually exceed $T_\mathrm{CMB}$. 

Before doing an in-depth analysis that includes astrophysical heating and imperfect WF coupling, it will be useful to determine which EMDE scenarios heat the IGM enough to have an observable impact on the global 21\nobreakdash-cm signal.
We forecast bounds on the DM annihilation rate assuming a detection of an absorption signal at $z = 17$. One motivation for the selection of $z=17$ is the 21\nobreakdash-cm absorption feature claimed by the EDGES collaboration, which was centered on $z\approx17$~\cite{Bowman2018}. 
As seen in Fig.~\ref{fig:TEvolution}, the energy injected into the IGM by DM annihilation can heat the gas above $T_K^{(0)}$ around this time. By $z = 17$, the spin temperature, which previously traced $T_\text{CMB}$, is approaching $T_K$ due to Ly$\alpha$ scattering. Thus, it is expected that $T_K\lesssim T_S <T_\text{CMB}$ at $z = 17$.  Given that the IGM is expected to be predominantly neutral at $z =17$,\footnote{Solving Eq.~(\ref{eq:x0}) leads to $x_e^{(0)}(z = 17) \approx 10^{-4}$. Even for $k_\text{cut}/k_\text{RH} = 40$, $x_e(z =17) \lesssim 10^{-3}$ for the range of $\langle \sigma v \rangle$ and $m_\chi$ we consider.}
Eq.~(\ref{eq:deltaT21Long}) can be translated into a bound on $T_K$ at $z = 17$ by setting $T_K \leq T_S$ and $x_\text{HI} = 1$:
\begin{equation}
    T_K(z = 17) \lesssim \left( 1- \frac{\delta T_{21}}{35 \text{ mK}} \right)^{-1} 49 \text{ K}.
    \label{eq:boundEQ}
\end{equation}
Thus, an observation of $\delta T_{21}$ at $z  = 17$ places an upper bound on $T_K$. Solving Eq.~(\ref{eq:T0}) sets the IGM temperature in the absence of energy injection as \mbox{$T_K^{(0)} = 6.9$~K} at \mbox{$z = 17$}. Unless there are exotic cooling mechanisms to further decrease $T_K$ below $T_K^{(0)}$~\cite{Barkana2018,Munoz2018,Hill2018,Falkowski2018} or additional radio background sources other than the CMB~\cite{Feng2018,Fialkov2019}, \mbox{$T_K^{(0)} = 6.9$~K} corresponds to a maximum possible absorption signal of \mbox{$\delta T_{21} = -220$~mK} at $z = 17$ via Eq.~(\ref{eq:boundEQ}).

In Section \ref{section:global}, we calculate the constraints on dark matter annihilation implied by $\delta T_{21} \leq -200$~mK at $z=17$, which is a slightly ($\sim\!\!10\%)$ weaker signal than the maximum absorption.
If $T_K = T_S$, this signal corresponds to $T_K = 7.3$~K from Eq.~(\ref{eq:boundEQ}) and does not allow much heating above $T_K^{(0)}$. We also include forecasts for $\delta T_{21} \leq -100$~mK, a signal that corresponds to a $\sim\!\! 100\%$ heating of the IGM due to DM ($T_K \leq 12.8$~K if $T_S = T_K$).  However, we note that Ly$\alpha$ scattering may not be efficient enough at $z=17$ for the spin temperature to reach the IGM temperature, as we will explore in Section \ref{section:Zeus_global}.  
If $T_S > T_K$ at $z=17$, these values of $T_K$ will generate larger values of $\delta T_{21}$, but we will see in Section \ref{section:Zeus_global} that the imperfect coupling between $T_S$ and $T_K$ does not affect the relative impact of DM annihilation set by these two thresholds.

\section{Results}\label{section:results}
\subsection{Forecasted constraints from the global 21\nobreakdash-cm signal at $z=17$}
\label{section:global}

We use \verb|DarkHistory| to compute the IGM temperature due to heating from DM annihilation in EMDE cosmologies and find the constraints on $m_\chi$ and $\langle \sigma v \rangle$ that follow from either $T_K \leq 12.8$~K or $T_K \leq 7.3$~K at $z = 17$ (to 0.01~K tolerance).  In this first analysis, we do not include any heating from astrophysical sources; we assume that all the energy that is injected into the IGM is due to annihilating DM to obtain conservative bounds on the annihilation rate.  In Section~\ref{section:Zeus_global}, we relax this assumption and consider how astrophysical heating affects and mimics the signatures of DM annihilation. 
\begin{figure*}
\includegraphics[width=\textwidth]{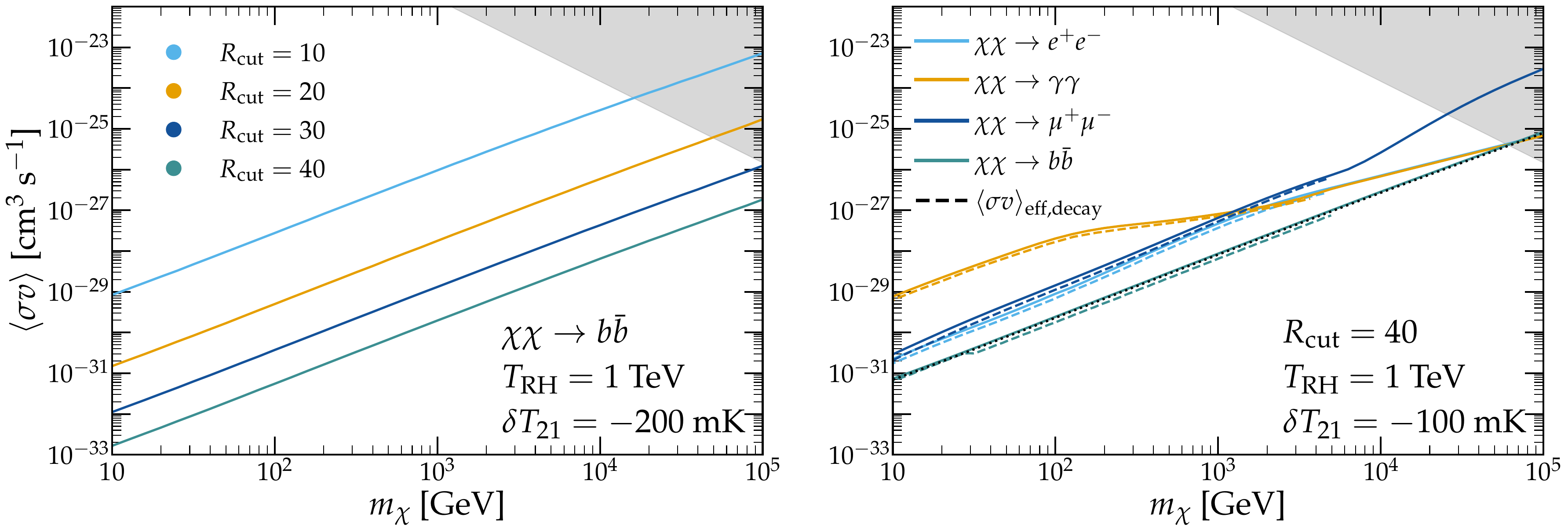}
\caption{The upper bounds on the velocity-averaged cross section for annihilating DM for various EMDE cosmologies and annihilation channels. 
Assuming that $T_K=T_S$, the left panel adopts $\delta T_{21}\leq-200$~mK ($T_K \leq 7.3$~K at $z = 17$), while the right panel adopts the weaker limit $\delta T_{21}\leq-100$~mK ($T_K \leq 12.8$~K at $z = 17$). In the left panel, we fix the $b\bar b$ annihilation channel but consider a range of $R_\mathrm{cut}$, while in the right panel, we fix $R_\mathrm{cut}=40$ but consider a range of channels. In both panels, the shaded region indicates that $\langle \sigma v \rangle > m_\chi^{-2}$, which violates unitarity.
The dashed lines in the right panel are the effective annihilation cross section bounds derived using Eq.~(\ref{eq:effectiveSigmav}) and the lower bounds on the lifetime of decaying DM from Ref.~\cite{Clark2018}. The concordance between the dashed and solid lines demonstrates how DM annihilation following an EMDE mimics decaying DM.}
\label{fig:constraints_rcuts} 
\end{figure*}

\begin{figure*}
\includegraphics[width=0.9\textwidth]{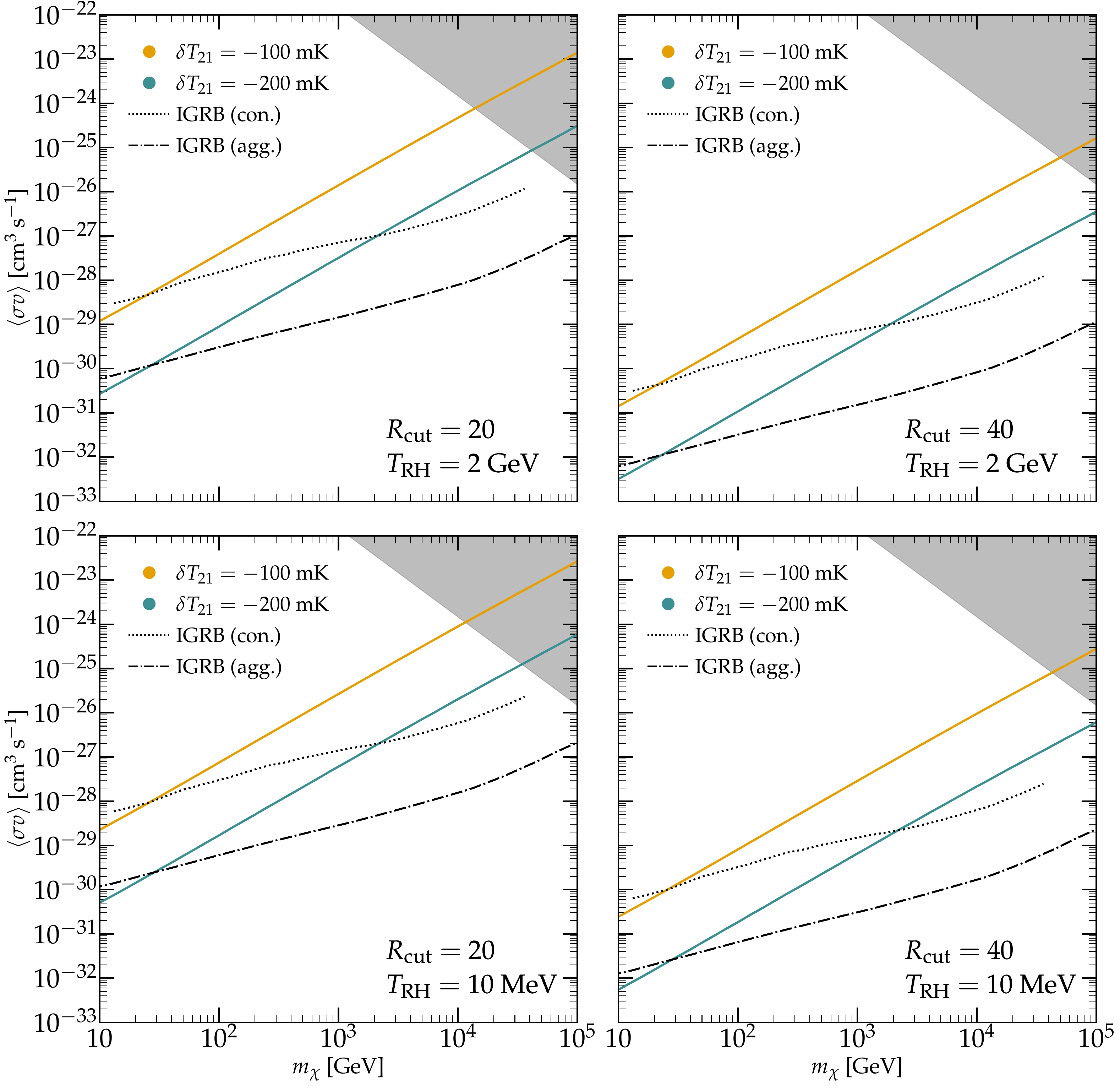}
\caption{Upper bounds on the cross section for DM annihilating into $b\bar{b}$ quarks assuming a detection of $\delta T_{21} =-100$~mK and $\delta T_{21} = -200$~mK at $z = 17$ with $T_S = T_K$ for different reheat temperatures and $R_\mathrm{cut}$ values.  Dotted and dash-dot lines are the conservative and aggressive bounds from the isotropic gamma-ray background (IGRB) from Ref.~\cite{Delos2019_DarkDegen}, respectively. The shaded region indicates that $\langle \sigma v \rangle > m_\chi^{-2}$, which violates unitarity.}
\label{fig:constraints} 
\end{figure*}

Figure~\ref{fig:constraints_rcuts} shows the predicted upper bounds on the cross section of annihilating DM. As discussed in section~\ref{section:Boost}, the ratio $R_\mathrm{cut} \equiv k_{\text{cut}}/k_{\text{RH}}$ sets the maximum enhancement of density perturbations. Consequently, larger values of $R_\mathrm{cut}$ correspond to more stringent bounds on $\langle \sigma v \rangle$, as seen in the left panel of Fig.~\ref{fig:constraints_rcuts}.  The right panel of Fig.~\ref{fig:constraints_rcuts} illustrates how the bounds on $\langle \sigma v \rangle$ depend on the annihilation channel. 
The upper bound on $\langle \sigma v \rangle$ for the photon annihilation channel becomes less sensitive to $m_\chi$ at energies above 100 GeV because pair production off the CMB becomes the dominant cooling mechanism at these energies, and the cross section for this interaction increases sharply with photon energy \cite{Slatyer2009}. The corresponding increase in the efficiency functions nearly compensates for the factor of $m_\chi^{-1}$ in the energy injection rate. This feature is not seen in the muon or quark channels because the photons produced by the decays of these annihilation products have broad spectra, which wash out energy-dependent features in the efficiency functions \cite{Clark2018}.

By comparing the $R_\mathrm{cut} = 40$, $\chi \chi \rightarrow b \bar{b}$ curves in the two panels of Fig.~\ref{fig:constraints_rcuts}, we see that decreasing the depth of absorption from $-200$~ mK to $-100$~ mK increases the upper limit on $\langle \sigma v \rangle$ by almost two orders of magnitude. 
Figure~\ref{fig:constraints} further illustrates the impact of changing the depth of 21\nobreakdash-cm absorption and also shows how the upper bounds of $\langle \sigma v \rangle$ are only weakly dependent on $T_\mathrm{RH}$.  Since increasing $T_\mathrm{RH}$ for a fixed value of $R_\mathrm{cut}$ only slightly increases the peak amplitude of the matter power spectrum, increasing $T_\mathrm{RH}$ leads to modestly stronger bounds on $\langle \sigma v \rangle$. 

The right panel of Fig.~\ref{fig:constraints_rcuts} demonstrates that our constraints are consistent with the limits on decaying DM from Ref.~\cite{Clark2018}, which also forecasts constraints for $T_K \leq 12.8$~K at $z=17$. As discussed in Section~\ref{section:EInj}, decaying DM gives the same energy injection rate per volume as annihilating DM after most of the microhalos have formed: both are proportional to the average DM density. If 21\nobreakdash-cm observations are predominantly sensitive to energy injection at $z<z_\mathrm{peak}$, the impact of DM annihilation within microhalos should be the same as the impact of a decaying DM particle that generates the same energy injection rate.  
For decaying DM with lifetime $\tau_\chi$,
\begin{equation}\label{eq:EinjDecay}
    \left( \frac{dE}{dVdt} \right)^{\text{inj}} = \langle \rho_{\chi,0} \rangle (1+z)^3 \frac{1}{\tau_\chi}.
\end{equation}
As seen from Fig.~\ref{fig:boost}, the boost factor becomes proportional to $(1+z)^{-3}$ long before a redshift of 17 in cosmologies with an EMDE. Thus, $\mathcal{B}(z) \approx \mathcal{B}(z = 0)(1+z)^{-3}$, and the energy injection rate due to annihilating DM can be approximated as 
\begin{equation}\label{eq:EinjApprox}
    \left( \frac{dE}{dVdt} \right)^{\text{inj}} \approx \langle \rho_{\chi,0} \rangle^2 (1+z)^3 \frac{\langle \sigma v \rangle}{m_\chi}  \mathcal{B}(z = 0).
\end{equation}
Before simply equating these two energy injection rates, we must consider the form the injection takes. If we assume that both annihilations and decays generate the same pair of particles, the decay of DM particles with mass $2m_\chi$ generates the same products as the annihilation of DM particles with mass $m_\chi$.  Therefore, we should equate Eq.~(\ref{eq:EinjDecay}) for DM particles with lifetime $\tau_\chi$ and mass $2m_\chi$ to Eq.~(\ref{eq:EinjApprox}): 
\begin{equation}\label{eq:effectiveSigmav}
    \langle \sigma v \rangle_\text{eff} = \frac{m_\chi}{\tau_\chi} 
    \frac{1}{\langle \rho_{\chi,0}\rangle \mathcal{B}(z=0)},
\end{equation}
where $\langle \sigma v \rangle_\text{eff}$ is the annihilation cross section for DM particles with mass $m_\chi$ that injects the same energy as decaying DM with lifetime of $\tau_\chi$ and a mass of $2 m_\chi$. 

We include the $\langle \sigma v \rangle_\text{eff}$ bounds obtained from the limits on the lifetime of decaying DM assuming \mbox{$\delta T_{21} = -100$~mK} and full Ly$\alpha$ coupling~\cite{Clark2018} in the right panel of Fig.~\ref{fig:constraints_rcuts}.  We see that the constraints derived from bounds on decaying DM are very similar to our constraints, which confirms that DM annihilation within microhalos has the same impact as decaying dark matter at $z=17$ for $R_\mathrm{cut} = 40$.  This is expected: for $R_\mathrm{cut} = 40$, ${\cal B}(z) \propto (1+z)^{-3}$ for $z\lesssim1000$, and the heating of the IGM at $z=17$ is not sensitive to the energy injection rate at redshifts greater than 1000. 

Since $B(z)\propto (1+z)^{-3}$ for all redshifts less than 100 even for $R_\mathrm{cut}=10$, as seen in Figure \ref{fig:boost}, 21\nobreakdash-cm observations cannot distinguish DM annihilation following an EMDE from decaying DM. If gamma-ray emission from dark matter is eventually detected, however, it may be possible to use the gamma-ray emission profiles of galaxies to differentiate these two scenarios \cite{Delos2019_DarkDegen} because microhalos are destroyed by stellar encounters \cite{Delos:2019tsl,Stucker2023} and tidal stripping \cite{Delos:2019lik} in the central regions of galaxies. For $R_\mathrm{cut} \simeq 10$, DM annihilation following an EMDE may also lead to a distinct signature in the CMB because there would be less energy injection at $z>100$ than would be produced by decaying DM with the same present-day energy injection rate.

The upper bounds on $\langle \sigma v \rangle$ for EMDE cosmologies from the IGRB shown in Fig.~\ref{fig:constraints} utilize the equivalence between microhalo-dominated annihilation and decaying DM~\cite{Delos2019_DarkDegen}. The IGRB measured by the \emph{Fermi} Collaboration \cite{Fermi-LAT:2014ryh} limits the lifetime of decaying DM, and this minimum lifetime was converted to an upper bound on $\langle \sigma v \rangle$ in EMDE cosmologies using Eq.~(\ref{eq:effectiveSigmav}). The conservative IGRB bound allows DM to generate the measured IGRB flux at each frequency~\cite{Liu2017}, whereas the aggressive bounds place stronger limits on the gamma-ray flux attributable to decaying DM by considering how astrophysical sources such as star-forming galaxies and active galactic nuclei contribute to the IGRB~\cite{Blanco2019_IGRB}.

Forecasts on the bounds of the annihilation cross section from $\delta T_{21}\leq -100$~mK ($T_K \leq 12.8$ K) at $z=17$ in cosmologies with an EMDE are stronger than the conservative IGRB bound for $m_\chi < 20$~GeV. Predicted constraints from $\delta T_{21}\leq -200$~mK ($T_K \leq 7.3$ K) at $z=17$ are stronger than the bounds from the aggressive IGRB bound for \mbox{$m_\chi < 20$~GeV}. These forecasted 21\nobreakdash-cm upper bounds on $\langle \sigma v \rangle$ are proportional to $m_\chi^{3/2}$; the relative inefficiency of heating by more massive particles seen in Fig.~\ref{fig:TEvolution} steepens the $m_\chi^{-1}$ scaling associated with the energy injection rate. 
Meanwhile, the upper bound on $\langle \sigma v \rangle_\text{eff}$ calculated from IGRB constraints on the dark matter lifetime is roughly proportional to $m_\chi^{3/4}$; the lower bound on $\tau_\chi$ is roughly proportional to $m_\chi^{1/4}$ for $m_\chi < 10^5$~GeV because the IGRB flux decreases with energy \cite{Blanco2019_IGRB}.
Hence, for smaller DM masses, a measurement of the global 21\nobreakdash-cm signal at a redshift of 17 could provide the most stringent bounds on $\langle \sigma v \rangle$ following an EMDE.

\subsection{Impact on the 21-cm global signal}
\label{section:Zeus_global}

Thus far, we have assumed that the spin and kinetic temperatures are coupled ($T_S = T_K$), in which case the 21\nobreakdash-cm signal is straightforwardly bounded by $\delta T_{21}$ as in Eq.~\eqref{eq:boundEQ}.
However, this complete WF coupling is not guaranteed. On the contrary, at early times in cosmic dawn, before X-ray sources turn on, the WF coupling need not fully couple $T_K$ and $T_S$~\cite{2017MNRAS.464.1365M,Park:2018ljd,Munoz:2021psm}.

We now lift the assumption of infinitely efficient WF coupling and use the 21\nobreakdash-cm model of Refs.~\cite{2023MNRAS.523.2587M,Cruz:2024fsv} to compute the properties of the 21\nobreakdash-cm background. 
We modified the publicly available code {\tt Zeus21}\footnote{\url{https://github.com/JulianBMunoz/Zeus21}} to take in $T_K(z)$ computed from \verb|DarkHistory| for our models. {\tt Zeus21} uses a galaxy formation model to compute the WF coupling and the X-ray flux from astrophysical sources and returns the updated kinetic temperature, the spin temperature, the 21\nobreakdash-cm global signal, and the power spectrum of fluctuations in $T_{21}$ as a function of redshift. 

We used {\tt Zeus21} to analyze the 21\nobreakdash-cm signals of DM particles with $m_\chi = 100\,\mathrm{GeV}$ and annihilation cross sections that match the upper limits for $\chi \bar \chi \rightarrow b\bar{b}$ derived in the previous section for an EMDE with $R_\mathrm{cut} = 20$ and a reheat temperature of 2 GeV. The cross sections for these two benchmark models are as follows:
\begin{itemize}
	\item{{\bf 100-mK bound}:$ \langle \sigma v \rangle = 3.95 \times 10^{-28} \, \mathrm{cm}^3\mathrm{s}^{-1}$;}
	\item{{\bf 200-mK bound}:$\langle \sigma v \rangle = 8.94 \times 10^{-30}\, \mathrm{cm}^3\mathrm{s}^{-1}$.}
\end{itemize}
These benchmark models are chosen, as described in the previous section, to give $\delta T_{21} = -100$~mK or $\delta T_{21} = -200$~mK at $z=17$ assuming perfect WF coupling.

\begin{figure}[btp!]
	\includegraphics[width=0.48\textwidth]{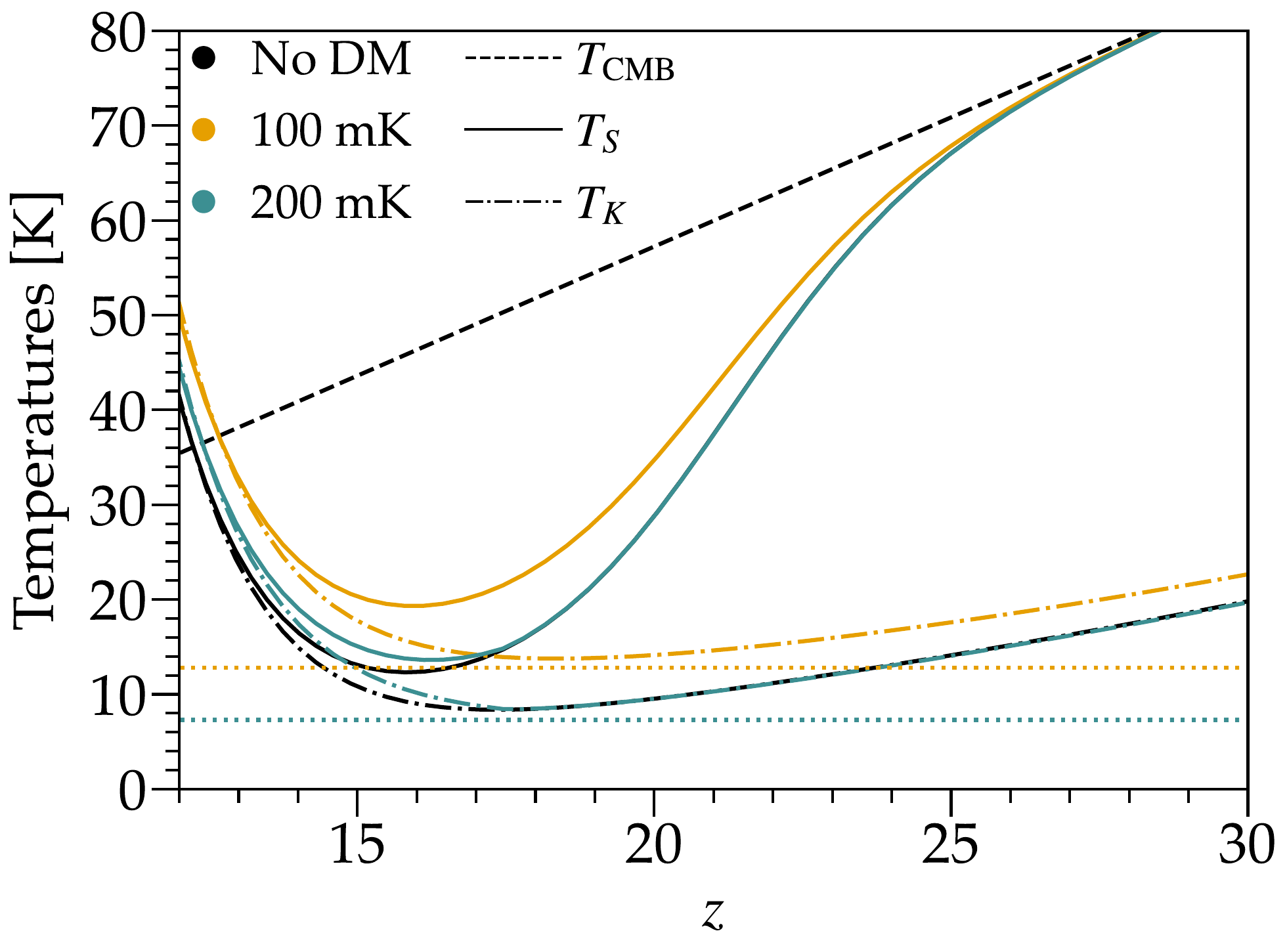}
	\caption{Evolution of different global temperatures against redshift $z$ as computed by {\tt Zeus21}. The dashed line shows the CMB temperature, which acts as the backlight for the 21-cm signal. The kinetic gas temperature $T_K$ and the spin temperature $T_S$ are shown for three thermal histories: the ``No DM" curves include no energy injection from DM, while the ``100~mK" and ``200~mK" curves show the effects of our two benchmark models. Both of these scenarios follow an EMDE with $T_\mathrm{RH} = 2$ GeV and $R_\mathrm{cut} = 20$ and assume a DM particle mass of 100 GeV that annihilates into $b\bar{b}$ quarks. The annihilation cross sections are chosen to lie slightly below the corresponding upper limits in Fig.~\ref{fig:constraints}. The horizontal dotted lines show the two limits on $T_K$ at $z=17$ adopted in Section \ref{section:global}. The $T_K$ curves shown here exceed these values due to the inclusion of X-ray heating from astrophysical sources. Furthermore, these curves show that $T_S \neq T_K$ at $z=17$. } 
\label{fig:tempevol} 
\end{figure}

The galaxy formation model of {\tt Zeus21} is fitted to the $4 \lesssim z \lesssim 10$ UV luminosity functions of the Hubble and James Webb Space Telescopes~\cite{Munoz:2023cup} and assumes that a fixed amount of Lyman-$\alpha$ and X-ray photons are emitted per unit star formation rate (SFR), following~Ref.~\cite{Mesinger:2010ne}. As a full astrophysical study of the 21\nobreakdash-cm signal is beyond the scope of this work, 
we refer the reader to Ref.~\cite{2023MNRAS.523.2587M} for a complete account of the free parameters within {\tt Zeus21}. The  most relevant parameter to our study is the X-ray efficiency $L_{40}$, defined as the luminosity emitted in the form of soft ($0.5-2$ keV) X-rays per unit SFR, in units of $10^{40}$~erg/s/($M_\odot$/yr); the default assumption is that $L_{40} = 3$, adopted from Ref.~\cite{Madau:2016jbv}.
Using this value for $L_{40}$ along with the other default parameters in {\tt Zeus21}, we show in Fig.~\ref{fig:tempevol} how $T_K$ and $T_S$ are predicted to evolve during cosmic dawn (i.e., $z\gtrsim 12$, before reionization is in full swing). For the 100\nobreakdash-mK benchmark model, DM annihilation has already significantly heated the IGM when the WF effect causes $T_S$ to pull away from $T_\mathrm{CMB}$ at $z\simeq27$, whereas the 200\nobreakdash-mK benchmark model only increases $T_K$ for $z\lesssim 17$. As seen in the ``No DM" case in Fig.~\ref{fig:tempevol}, X-ray heating from astrophysical sources causes $T_K$ to increase with time for $z\lesssim 16$. Figure~\ref{fig:tempevol} also shows that WF coupling is not yet complete when $T_S$ reaches its minimum value: $T_S > T_K$ for $z\gtrsim 13$.
At lower redshifts, we see $T_K$ and $T_S$ converging, but by then there is significant X-ray heating from astrophysical sources in our fiducial galaxy formation model.

\begin{figure}
\includegraphics[width=0.48\textwidth]{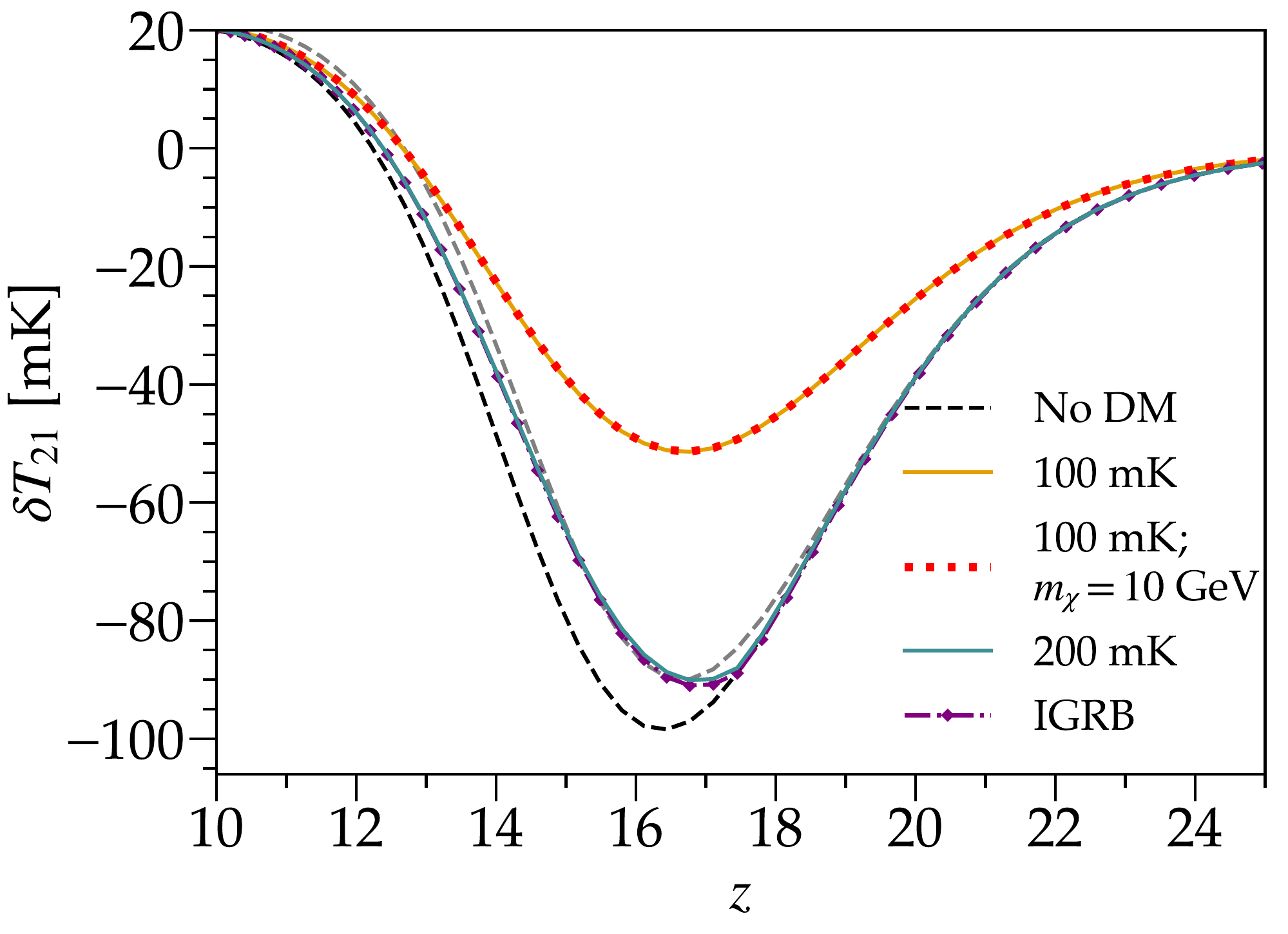}
	\caption{21-cm global signal against redshift for our different models showing the expected absorption trough around \mbox{$z\sim 17$}. The ``No DM" scenario has the deepest absorption, as the gas is coldest. Adding DM annihilation heats the gas, leading to a shallower trough. In addition to the 100-mK and 200-mK benchmark scenarios shown in Fig.~\ref{fig:tempevol}, we show the global signal for two additional scenarios that also follow an EMDE with $T_\mathrm{RH} = 2$ GeV and $R_\mathrm{cut} = 20$. The red-dotted curve has $m_\chi = 10$ GeV and $\langle \sigma v \rangle = 1.19\times 10^{-29} \, \mathrm{cm}^3\mathrm{s}^{-1}$. This particle also falls near the 100-mK constraint in Fig.~\ref{fig:constraints}, and its 21-cm global signal is indistinguishable from the 100-mK benchmark. The dot-dashed IGRB curve shows $m_\chi = 100$~GeV and $\langle \sigma v \rangle = 3.06 \times 10^{-30}  \, \mathrm{cm}^3\mathrm{s}^{-1}$, which falls on the aggressive IGRB constraint in Fig.~\ref{fig:constraints}; its global signal barely differs from the 200-mK benchmark scenario even though the annihilation cross section is one-third as large. The heating from the IGRB and 200-mK models can be mimicked in the global signal by increasing the luminosity of astrophysical X-rays by 33\%, as shown by the gray dashed line.
    }
	\label{fig:global} 
\end{figure}

The impact of imperfect WF coupling is apparent in the 21\nobreakdash-cm global signal shown in Fig.~\ref{fig:global}.
The standard (no DM) case has an absorption trough that peaks at $z\sim 17$, reaching a depth of $\delta T_{21}\approx -100$ mK, as opposed to the $\delta T_{21}\approx -220$ mK trough that would result from perfect WF coupling and no astrophysical heating. 
Nevertheless, we see that adding DM heating decreases the depth of the absorption.
In particular, the 100\nobreakdash-mK benchmark model reduces the signal by a factor of two, to $\delta T_{21} \sim -50$ mK (independently of the DM mass, as shown by the nearly identical case with $m_\chi = 10\, \mathrm{GeV}$ and $\langle \sigma v \rangle = 1.19 \times 10^{-29}\, \mathrm{cm}^3\mathrm{s}^{-1}$ in Fig.~\ref{fig:global}).
A more optimistic reach is represented by the 200\nobreakdash-mK benchmark model, which remains close to the no-DM case at $T_{21}\approx -90$~mK, similar to the signal that saturates the aggressive IGRB bound for $m_\chi = 100$ GeV, $T_\mathrm{RH} = 2$ GeV, and $R_\mathrm{cut} = 20$: $\langle \sigma v \rangle < 3.1 \times 10^{-30}\, \mathrm{cm}^3\mathrm{s}^{-1}$.
In summary, although imperfect WF coupling alters the depth of 21\nobreakdash-cm absorption troughs for our benchmark DM heating scenarios, their relative impact is not affected. Therefore, the ``$\delta T_{21} = -100$~mK" and ``$\delta T_{21} = -200$~mK" constraints presented in Section~\ref{section:global} denote the parameters that generate a 50\% and 10\% reduction in trough depth, respectively.

One of the main systematics that may prevent us from detecting DM heating is its degeneracy with astrophysical X-rays~\cite{Sun:2023acy,Facchinetti:2023slb}.
We illustrate this in Fig.~\ref{fig:global}, where the gray dashed curve shows the 21\nobreakdash-cm signal from a scenario without DM energy injection but 33\% more astrophysical X-ray emission ($L_{40}=4$ instead of the default $L_{40}=3$). 
The impact of this additional X-ray heating is very similar to the impact of DM annihilation for the 200\nobreakdash-mK benchmark model. In contrast, the global signal for the 100\nobreakdash-mK benchmark model cannot be mimicked by increasing the X-ray luminosity. Reducing the depth of the trough to 50~mK requires dramatically increasing the X-ray luminosity to $L_{40}=16$, which leads to an emission signal ($\delta T_{21}>0$) for $z\lesssim 15$. We conclude that the 21\nobreakdash-cm global signal could be used to detect DM annihilation at the level of the 100\nobreakdash-mK model, but DM annihilation at the level of the 200\nobreakdash-mK model would be difficult to detect even if we had a perfect measurement of the global signal due to uncertainties in the X-ray luminosity. 
Fortunately, there is information beyond the global signal. 

\subsection{Impact on the 21-cm power spectrum}
\label{section:Zeus_ps}

There are several experimental efforts to detect $\delta T_{21}$ through its power spectrum, which is more robust to foregrounds as these are confined to a ``wedge'' in Fourier space~\cite{Parsons:2012qh,Liu:2014yxa}.
Let us briefly investigate how the 21\nobreakdash-cm power spectrum changes when adding energy injection from annihilating DM following an EMDE. 
The same {\tt Zeus21} code we used above can also return the 21\nobreakdash-cm power spectrum under the assumption that the heating of the IGM is inhomogeneous like the heating from astrophysical X-rays or entirely homogeneous. 
The DM-induced heating we consider here is neither, but on kiloparsec scales, its spatial fluctuations arise from the clustering of the microhalos as opposed to the microhalos themselves.  Since the microhalos track the DM density, the heating from DM annihilation within microhalos has nearly the same spatial variation as the emission from decaying DM.\footnote{The emission profile within galaxies for microhalo-dominated annihilation can be suppressed in the central region relative to the emission profile for decaying DM due to the disruption of microhalos by tidal forces and stellar encounters~\cite{Delos2019_DarkDegen}. This effect would make heating by annihilation within microhalos slightly more homogeneous than heating by decaying DM.}
The spatial inhomogeneity of decaying dark matter does not significantly affect its impact on the 21\nobreakdash-cm power spectrum \cite{Sun:2023acy}, so we treat microhalo-dominated annihilation as a homogeneous heating source. 
We also note that in this illustrative analysis, we have only included galaxies above the atomic-cooling threshold as in Ref.~\cite{2023MNRAS.523.2587M}, whereas X-rays from low-metallicity low-mass galaxies could change the picture~\cite{Cruz:2024fsv,Lazare:2023jkg}. Finally, we continue to keep the astrophysical parameters in {\tt Zeus21} fixed at their fiducial values when including DM annihilation. As in the previous section, we also consider the impact of increasing the X-ray luminosity in the absence of DM annihilation.

 \begin{figure}
\includegraphics[width=0.48\textwidth]{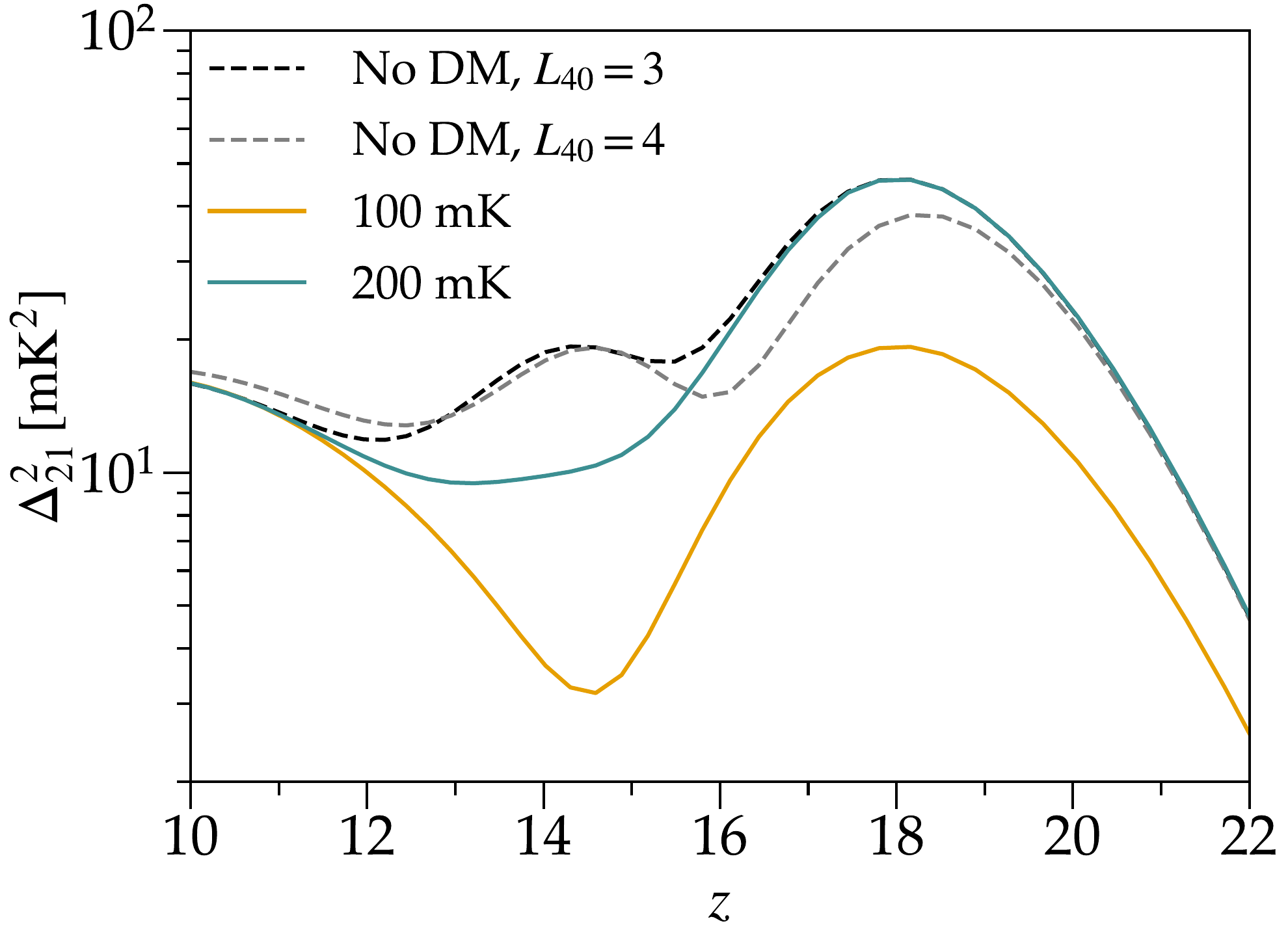}
\caption{The 21-cm power spectrum against redshift at a fixed scale $k=0.2$ Mpc$^{-1}$ for the 100-mK and 200-mK benchmark scenarios and the two scenarios with no DM annihilation shown in Fig.~\ref{fig:global}. In the absence of DM annihilation (dashed lines), there are two bumps in $\Delta_{21}^2(z)$: inhomogeneities in the efficiency of WF coupling enhance $\Delta_{21}^2$ at $z\sim 18$, and X-ray heating enhances $\Delta_{21}^2$ at $z\sim 14$ due to the clustering of heating sources (galaxies).  The 100-mK DM scenario illustrates how heating the gas reduces the 21-cm power; in this scenario, DM annihilation increases $T_K$ throughout the redshift range shown here, as seen in Fig.~\ref{fig:tempevol}. More importantly, the source of the heating can be teased out from the power spectrum.  The bump at $z\sim14$ is not present if the gas is heated by annihilation within DM microhalos because the heating is assumed to be homogeneous. Therefore, the 21-cm power spectrum could be used to differentiate between astrophysical and DM heating scenarios that generate similar global signals. 
}
\label{fig:21cmpowerspectrum} 
\end{figure} 

Under these assumptions, we show in Fig.~\ref{fig:21cmpowerspectrum} the evolution of the 21\nobreakdash-cm power spectrum $\Delta^2_{21}$ at a fixed scale of $k=0.2$ Mpc$^{-1}$. We stop our calculation at $z\sim 10$, where we expect reionization to begin in full force~\cite{Robertson:2015uda,Munoz:2024fas}, as {\tt Zeus21} does not yet model inhomogeneities in the ionized fraction.
The no-DM case shows the standard two peaks in $\Delta^2_{21}(z>10)$~\cite{Munoz:2021psm}: the peak at $z\sim 18$ arises from spatial fluctuations in the WF coupling, and the peak at $z\sim14$ is attributable to the inhomogeneity of X-ray heating, with a dip in between where the two effects move the signal in opposite directions. 

Additional IGM heating leads to an overall suppression of the 21\nobreakdash-cm power spectrum that roughly corresponds to its impact on the 21\nobreakdash-cm global signal. For example, the 100\nobreakdash-mK benchmark model reduces the depth of the global absorption trough by a factor of two and generates a similar suppression in $\Delta^2_{21}$ for all $z\gtrsim 15$. 
However, the 200\nobreakdash-mK benchmark model and the ``No DM" model with 33\% more X-ray heating shown in Fig.~\ref{fig:global} do not uniformly suppress $\Delta^2_{21}$ at high redshifts. Instead, increasing the X-ray luminosity only suppresses the WF peak at $z\sim18$. (This peak is also sensitive to changes to the star formation rate.) In contrast, the 200\nobreakdash-mK benchmark model only suppresses the X-ray heating peak at $z\sim14$. For $m_\chi = 100$~GeV, the annihilation cross section that saturates the IGRB bound on DM annihilations in microhalos generates nearly the same 21\nobreakdash-cm power spectrum as the 200\nobreakdash-mK benchmark model, as expected from the similarity between their global signals seen in Fig.~\ref{fig:global}. 

The differences in $\Delta^2_{21}$ for models with similar global signals but different heating sources provide a promising avenue for detecting DM annihilations and decays.  Our 200\nobreakdash-mK benchmark model leads to a roughly 10\% change in the global signal that can be mimicked by increasing the X-ray luminosity by 33\% as seen in Fig.~\ref{fig:global}. However, Fig.~\ref{fig:21cmpowerspectrum} shows that the 200\nobreakdash-mK model halves $\Delta^2_{21}$ for $k=0.2$ Mpc$^{-1}$ and $z\sim 14$ compared to both models that do not include heating from DM.
This example illustrates that even a modest amount of DM heating erases the X-ray bump at $z\sim 14$; DM energy injection is far more homogeneous than heating from astrophysical sources and thus does not produce a peak in the power spectrum. Similar results have been reported for decaying DM in Refs.~\cite{Evoli:2014pva, Lopez-Honorez:2016sur, Sun:2023acy, Facchinetti:2023slb}. 

Reducing $\langle \sigma v \rangle$ to the aggressive IGRB limit generates nearly the same 21\nobreakdash-cm power spectrum as our 200\nobreakdash-mK benchmark model for $m_\chi \simeq 100$ GeV. Therefore, an ${\cal O}(1)$ detection of $\Delta^2_{21}$ at $z\sim 14$ could significantly improve constraints on DM annihilation following an EMDE for particles with $m_\chi \lesssim 100$ GeV. 
Moreover, combining measurements of the sky-averaged $\delta T_{21}$ with measurements of its power spectrum at $z\sim 14$ could provide evidence for DM heating that cannot be mimicked by increasing the X-ray luminosity of astrophysical sources. 
This target may be beyond the reach of HERA, as thermal noise increases steeply toward higher $z$, where foregrounds are stronger. As a result, the anticipated limits on the DM lifetime based on 1000 hours of HERA observations only exceed the constraints from X-ray and gamma-ray observations for $m_\chi \lesssim$~GeV~\cite{Facchinetti:2023slb}. 
However, the SKA may be able to provide the measurement of $\Delta^2_{21}$ required to improve upon the EMDE annihilation bounds from the IGRB for $m_\chi \lesssim 100$~GeV. Although its thermal noise also increases as it looks to higher redshift, forecasts predict  that the SKA should be able to detect $\Delta^2_{21} \gtrsim 10 \,\mathrm{mK}^2$ for $k=0.2$~Mpc$^{-1}$ and $z\lesssim 16$~\cite{Koopmans2015}. Such a detection would improve upon IGRB constraints on both the DM lifetime and DM annihilation following an EMDE, while the absence of this signal would indicate that the sources responsible for heating of the IGM were homogeneous on 100-kiloparsec scales. 

\section{Conclusion}\label{section:conclusion}

The linear growth of DM density perturbations during an EMDE leads to sub-Earth-mass microhalos that are more dense and form earlier than halos in standard cosmology. Dark matter particles with annihilation rates boosted by these microhalos inject energy into the IGM, increasing the IGM temperature $T_K$. Hence, an upper bound on the IGM temperature would limit the DM annihilation cross section $\langle \sigma v \rangle$ for a DM particle with mass $m_\chi$. After the first stars form around $z\simeq 20$, efficient scattering of Ly$\alpha$ photons is expected to couple the spin temperature $T_S$ of the 21\nobreakdash-cm signal to the IGM kinetic temperature, and we explore the potential of 21\nobreakdash-cm observations as probes of the DM annihilation cross section in EMDE cosmologies.

We use the Peak-to-Halo (P2H) method~\cite{Delos2019_PredictDens,Delos2019_DarkDegen} to calculate the DM annihilation rate in EMDE cosmologies with reheat temperatures of $T_\text{RH} =$ 10~MeV, 1~GeV, and 1~TeV, and cut ratios $R_\mathrm{cut} \equiv k_\text{cut}/k_\text{RH} = $10, 20, 30, and 40. The boost to the annihilation rate from microhalos that form after an EMDE can be several orders of magnitude larger than the boost expected from standard structure formation, with $\mathcal{B}(z=0) > 10^{11}$ for $k_\text{cut}/k_\text{RH} = 40$. Furthermore, $\mathcal{B}(z)$ becomes significant at higher redshifts in cosmologies with an EMDE, and the energy injected by annihilating DM can dramatically alter the evolution of the IGM temperature at $z>10$.

If no energy is injected into the IGM, it is expected that $T_K = 6.9$~K at $z = 17$, corresponding to a 21\nobreakdash-cm signal of $\delta T_{21} \approx -220$~mK if $T_K = T_S$. We consider $\delta T_{21}~\leq~-200$~mK and $\delta T_{21}~\leq~-100$~mK at $z = 17$ to forecast bounds on $\langle \sigma v \rangle$ in EMDE cosmologies with the assumption that DM annihilation is the only source of energy injection. A signal of $\delta T_{21}~\leq~-200$~mK at this redshift corresponds to a maximum IGM temperature of $T_K \leq 7.3$~K ($\sim\!\!10\%$ heating). Such a measurement would strongly limit IGM heating and would probe DM annihilation cross sections as small as $\langle \sigma v \rangle \approx 10^{-32}$~cm$^3$s$^{-1}$ for $m_\chi \simeq 10$ GeV following an EMDE with $R_\mathrm{cut} = 40$. A less restricting possibility is a signal of $\delta T_{21} \leq -100$~mK, corresponding to $T_K \leq 12.8$~K ($\sim\!\! 100\%$ heating). The resulting forecast of the upper bound on $\langle \sigma v \rangle$ can be as small as $10^{-31}$~cm$^3$s$^{-1}$ for the same parameters.  Since DM freeze-out during or before an EMDE can only generate the observed DM abundance for $\langle \sigma v \rangle \ll 10^{-26}$~cm$^3$s$^{-1}$ \cite{Erickcek2015, Delos2019_DarkDegen}, such stringent limits on the annihilation cross section are required to constrain DM production in EMDE cosmologies \cite{Blanco2019, Ganjoo:2024hpn}.

We limit our calculations to $k_{\text{cut}}/k_{\text{RH}} \leq 40$, as the P2H method is only valid for halos that form during matter domination. The majority of microhalos form after $z\approx3000$ for $k_{\text{cut}}/k_{\text{RH}} = 40$, but larger $k_{\text{cut}}/k_{\text{RH}}$ values would require more careful modeling of halo formation during radiation domination~\cite{Blanco2019, Ganjoo:2023fgg}. We expect that larger values of $k_{\text{cut}}/k_{\text{RH}}$ would lead to larger $\mathcal{B}(z)$ and thus stronger constraints on the annihilation cross section. 

Existing constraints on $\langle \sigma v \rangle$ in EMDE cosmologies \cite{Delos2019_DarkDegen, Blanco2019, Ganjoo:2024hpn} are derived from Fermi-LAT observations of the isotropic gamma-ray background (IGRB)~\cite{Fermi-LAT:2014ryh}. These analyses used lower bounds of the lifetime of decaying DM from the IGRB \cite{Liu2017, Blanco2019_IGRB} to deduce
upper bounds of the annihilation cross section in EMDE cosmologies; like decaying DM, annihilating DM within microhalos injects a constant energy per DM mass after the microhalo population is established. Conservative limits on the DM lifetime are derived by requiring the gamma-ray flux due to decaying DM to be less than the flux reported by Fermi-LAT~\cite{Liu2017}. Accounting for astrophysical contributions to the IGRB leads to more aggressive limits~\cite{Blanco2019_IGRB}.

We follow Ref.~\cite{Delos2019_DarkDegen} in assuming that the $\rho \propto r^{-3/2}$ inner density profiles soften into the NFW profile over time. However, recent simulations have shown that the $\rho \propto r^{-3/2}$ cusps survive~\cite{Delos2022_cuspSurvival,Delos:2025pen}.
The annihilation rate from these ``prompt cusps'' greatly exceeds that from NFW halos (e.g.~\cite{Delos2022_cuspAnnihilationEffect, Delos:2023vfv,Delos:2023ipo,Crnogorcevic:2025nwp}), and as discussed in Section~\ref{section:Boost}, accounting for their survival would increase the annihilation boost factor by about a factor of 5. The upper bounds on $\langle \sigma v \rangle$ would then decrease by the same factor. Note, however, that the annihilation rate in prompt cusps depends on details of the cosmological scenario beyond our simple $(T_\text{RH},R_\mathrm{cut})$ parametrization, since it is influenced not only by the spatial distribution of the dark matter after the EMDE but also by its velocity distribution. These considerations were included in a recent analysis of IGRB constraints on DM annihilation following an EMDE \cite{Ganjoo:2024hpn}. We also note that, as discussed in Refs.~\cite{Delos2022_cuspAnnihilationEffect,Delos:2023ipo,Crnogorcevic:2025nwp,Ganjoo:2024hpn}, the annihilation rate from the prompt cusps has systematic uncertainty amounting to about a factor of 2, which arises from a combination of uncertainty in the numerical factor in Eq.~(\ref{eq:A}) and uncertainty in the fraction of initial peaks that can be associated with surviving cusps at late times. Since the primary objective of this work is to evaluate the potential of 21\nobreakdash-cm observations as probes of EMDE cosmologies, we make the same assumptions as Ref.~\cite{Delos2019_DarkDegen} so that we can directly compare limits on $\langle \sigma v \rangle$. However, using a more complete description of the EMDE scenario (as in Ref.~\cite{Ganjoo:2024hpn}), future work could improve on the $\langle \sigma v \rangle$ limits in this work by a significant factor.

For $m_\chi \lesssim 20$~GeV, requiring $\delta T_{21} \leq -100$~mK leads to more stringent constraints than those derived from the conservative IGRB limit on the DM lifetime. Requiring $\delta T_{21}~\leq~-200$~mK leads to stronger bounds than those from the aggressive IGRB for $m_\chi \lesssim 20$~GeV. We expect this trend to continue at lower masses. \verb|DarkHistory| utilizes electron and photon spectra from DM with masses between 5~GeV and 100~TeV annihilating into the SM for DM particles~\cite{Cirelli2011}. Our constraints on $\langle \sigma v \rangle$ could be extended to lower DM masses with a broader range of available injection spectra. However, if $m_\chi \lesssim 0.1 T_\mathrm{RH}$, the DM will freeze out after the EMDE, in which case the EMDE will not affect its small-scale structure or its annihilation rate. Therefore, observations of the global 21\nobreakdash-cm signal are unlikely to improve upon the IGRB constraints on EMDE cosmologies with $T_\mathrm{RH} \gtrsim 2$~GeV.

While small changes to the IGM temperature generate similarly small changes to the global 21\nobreakdash-cm signal, energy injection from DM annihilation within microhalos has an amplified effect on the spatial inhomogeneity of the 21\nobreakdash-cm brightness temperature at $z\simeq 14$. If astrophysical sources are responsible for heating the IGM, the 21\nobreakdash-cm power spectrum is amplified at $z\sim 14$ due to the inhomogeneity introduced by the first X-ray sources.  In contrast, the heating from DM annihilation within microhalos tracks the DM density and can be approximated as homogeneous \cite{Sun:2023acy}.  As a result, there is no enhancement to the 21\nobreakdash-cm power spectrum at $z\sim 14$, and the $\sim\!\!10\%$ change in the IGM temperature that corresponds to our $\delta T_{21} \leq -200$~mK benchmark reduces the 21\nobreakdash-cm power at $k = 0.2$~Mpc$^{-1}$ by a factor of two.  For DM particle masses less than 100 GeV, EMDE models that saturate bounds on DM annihilation from the IGRB have nearly the same impact on the 21\nobreakdash-cm power spectrum as this benchmark model, so a measurement of the 21\nobreakdash-cm power spectrum at $z\sim 14$ could provide powerful new constraints on DM annihilation following an EMDE. 
Therefore, the abundance of ongoing and planned experimental efforts to measure the 21\nobreakdash-cm power spectrum will provide a promising avenue to understand the expansion history prior to big bang nucleosynthesis and the origins of dark matter.

\acknowledgements
We thank Hector Cruz for useful discussions and assistance with {\tt Zeus21}.
HB and ALE were supported by NSF Grants No. PHY-1752752 and PHY-2310719.
JBM was supported at UT Austin by NSF Grants No. AST-2307354 and AST-2408637.
The authors thank the Kavli Institute for Theoretical Physics (KITP) for their hospitality during part of this work. This research was supported in part by grant NSF PHY-2309135 to the KITP.

\end{document}